\documentclass[
 aip,
 jcp,
 amsmath,amssymb,
 reprint
]{revtex4-1}

\usepackage[utf8]{inputenc}
\usepackage[T1]{fontenc}
\usepackage[version=3]{mhchem} 
\usepackage{xcolor}
\usepackage{bm}
\usepackage{amsmath}
\usepackage{amssymb}
\usepackage{txfonts}
\usepackage{graphicx}
\usepackage{xspace}
\usepackage{float}
\usepackage{units}
\usepackage{color}
\usepackage{booktabs}
\usepackage{multirow}
\usepackage{gensymb}
\usepackage{hyperref}
\usepackage{ulem}
\usepackage{mathptmx}
\usepackage{etoolbox}

\makeatletter
\def\@email#1#2{%
 \endgroup
 \patchcmd{\titleblock@produce}
  {\frontmatter@RRAPformat}
  {\frontmatter@RRAPformat{\produce@RRAP{*#1\href{mailto:#2}{#2}}}\frontmatter@RRAPformat}
  {}{}
}%
\makeatother

\begin{document}

\title[]{Deep Potential-Driven Molecular Dynamics of CO Ice Analogues : Investigating Desorption Following Vibrational Excitation}

\author{Maxime Infuso}
\affiliation{Univ. Lille, CNRS, UMR 8523 – PhLAM – Physique des Lasers Atomes et Molécules, F-59000 Lille, France}
\email{maxime.infuso@univ-lille.fr}

\author{Samuel Del Fré}
\altaffiliation[Also at ]{Univ. Lille, CNRS, INRAE, Centrale Lille, UMR 8207 - UMET - Unité Matériaux et Transformations, F-59000 Lille, France}
\email{samuel.del-fre@univ-lille.fr} 

\author{Gilberto A. Alou}
\affiliation{Univ. Lille, CNRS, UMR 8523 – PhLAM – Physique des Lasers Atomes et Molécules, F-59000 Lille, France}
\email{gilberto.alou@univ-lille.fr} 

\author{Mathieu Bertin}
\affiliation{Sorbonne Université, CNRS UMR 8233, De la Molécule aux Nano-Objets: Réactivité, Interactions et Spectroscopies, MONARIS, 75005, Paris, France.}
\email{mathieu.bertin@sorbonne-universite.fr}

\author{Jean-Hugues Fillion}
\affiliation{Sorbonne Université, CNRS UMR 8233, De la Molécule aux Nano-Objets: Réactivité, Interactions et Spectroscopies, MONARIS, 75005, Paris, France.}
\email{jean-hugues.fillion@sorbonne-universite.fr}

\author{Alejandro Rivero Santamar\'{i}a}
\affiliation{Univ. Lille, CNRS, UMR 8523 – PhLAM – Physique des Lasers Atomes et Molécules, F-59000 Lille, France}
\email{alejandro.rivero@univ-lille.fr}

\author{Maurice Monnerville}
\affiliation{Univ. Lille, CNRS, UMR 8523 – PhLAM – Physique des Lasers Atomes et Molécules, F-59000 Lille, France}
\email{maurice.monnerville@univ-lille.fr}


\date{\today}

\begin{abstract}
  We present a new deep learning-based machine learning potential (MLP) for molecular dynamics simulations of solid carbon monoxide (CO), capable of accurately describing CO vibrations both in the fundamental state and in highly excited vibrational states, up to approximately $v = 40$. The MLP is based on the combination of high-dimensional neural network atomic potentials using the DeePMD-kit package, trained on prior \textit{ab initio} molecular dynamics (AIMD) data, with selective treatment of the excited molecule allowing us to capture complex energy redistribution dynamics in condensed-phase environments. In particular, the MLP is capable of accurately describing the desorption process of a single CO molecule within an aggregate of 50 CO molecules, in excellent agreement with both previous theoretical predictions and experimental measurements. The MLP provides a much finer description of the translational and rotational energy distributions, capturing their character with high fidelity and allowing a more detailed comparison with experimental results. Furthermore, the analysis of the rotational energy, resolved over specific translational energies, revealed new insights into the coupling between translational and rotational degrees of freedom during the photodesorption process. This novel approach opens new perspectives for extensive statistical studies on desorption energies and detailed investigations of surface molecule excitations and the exploration of larger-scale models incorporating periodic boundary conditions to simulate more realistic CO aggregates.
\end{abstract}

\maketitle


\section{Introduction}

In the cold region of the interstellar medium (ISM), the freezing-out of atomic and molecular species onto dust particles leads to the formation of icy mantles playing the role of molecular reservoirs in these regions\cite{gibbInventoryInterstellarIces2000,boogert2015,mcclureIceAgeJWST2023}. 
These ices constantly interact with their environment, particularly with VUV photons originating for instance, from the interstellar radiation field, the central star in a protoplanetary disk or in the denser regions, generated as secondary photons from the interaction between cosmic rays and gaseous H$_2$ \cite{caselliOurAstrochemicalHeritage2012,prasadUVRadiationField1983,shenCosmicRayInduced2004}. Consequently, photodesorption at the ice surface — which can enrich the gas phase even below sublimation temperatures — is considered one of the major factors explaining anomalous gas-phase molecular abundances in the cold ISM, as is the case, for example, with carbon monoxide (CO) (eg., \cite{willacy2000,pietu2007,dartoisStructureDMTau}).
CO is indeed ubiquitous in the ISM and plays a pivotal role in the physics and chemistry of various regions of the universe \cite{dishoeckAbundanceInterstellarCO1987,obergPhotochemistryAstrochemistryPhotochemical2016a}. Therefore, its VUV photodesorption for different photons energy and sources, temperatures and ice morphology and composition has been extensively studied in laboratory \cite{oberg2007,munoz2010,fayolle2011,bertin2012,bertin2013,paardekooper2016,carrascosaIntriguingBehaviorUltraviolet2021,gonzalezdiazAccretionPhotodesorptionCO2019,oberg2009,sie2022,chenVACUUMULTRAVIOLETEMISSION2013,munoz2016}. These studies have shown that the photodesorption from pure CO mostly occurs via a process referred to as an indirect DIET (Desorption Induced by Electronic Transition) process, wherein the electronic excitation of a solid CO molecule within the 2 to 5 topmost molecular layers results in the desorption of another, surface-located one. In the 7 - 10.5 eV energy range, the photodesorption is solely triggered by the $A^1\Pi-X^1\Sigma^+$ electronic transition in the solid CO \cite{luSpectraVacuumUltraviolet2005, fayolle2011}. These experimental studies could not, however, explain by which mechanism energy transfer between the excited and the desorbing molecule takes place. Indeed, a comprehensive understanding of the molecular-level energy transfers involved in the desorption process requires sophisticated modeling due to the high complexity of the DIET mechanism. This task has recently been tackled in our prior studies \cite{delfreMechanismUltravioletInducedCO2023,hacquardPhotodesorptionCOIces2024} in which the final part of the DIET mechanism, that corresponds to the vibrational relaxation of one molecule from a highly excited vibrational state, was simulated using \textit{ab initio} molecular dynamics (AIMD) simulations based on Density Functional Theory (DFT). The results were compared with new, nanosecond laser-based experimental measurements that give access to the rovibrational state and kinetic energy of the desorbates. The theoretical results showed that CO molecules can effectively desorb following a three steps mechanism leading to the ejection of vibrationnally cold CO (96\% of $v = 0$) with relatively low translational ($< 200~\text{meV}$) and rotational energy (J $< 30$) for the large majority of desorbates, despite the initial vibrational excitation energy ($v = 40$, 8.26 eV). The theoretical energy distributions of the desorbed species match the experimental data remarkably well, highlighting the critical role of vibrational relaxation in the desorption process.
However, the high computational cost of AIMD necessitated limiting the system size (to 50 molecules) as well as the number and length of simulations (to 100 simulations and 5~ps, respectively). Although the current statistics are very satisfactory, extending the simulations to a larger scale could further reinforce the established findings while incorporating other critical aspects of the mechanism. 
To address these computational cost limitations, the development of high-dimensional machine learning potentials (MLP), trained on DFT data obtained on-the-fly via AIMD simulations, has become increasingly popular in recent years as these potentials are specifically designed to facilitate larger-scale simulations \cite{tokitaHowTrainNeural2023, unkeMachineLearningForce2021, wangMachineLearningInteratomic2024}. This approach enables classical MD simulations to be performed at a fraction of the cost of AIMD with full-dimensional potential energy surface (PES) while preserving DFT-level accuracy (eg., \cite{riverosantamariaHighDimensionalAtomisticNeural2021,chenAENETLAMMPSAENET2021}). 
The development of a MLP capable of simultaneously describing condensed-phase CO near its equilibrium geometry and in highly vibrationally excited state is particularly challenging. 
Since the condensed-phase CO PES exhibits a shallow energy topography, the neural network fit may incorporate minor fluctuations that are nevertheless significant enough to destabilize molecular dynamics simulations, thus necessitating precise and rigorous tuning of the training conditions. In addition, interactions among CO molecules near their equilibrium state are significantly weaker than those involving a highly vibrationally excited CO. This may introduce disparities in the training set energies and forces, potentially leading to extrapolation issues \cite{tokitaHowTrainNeural2023}.
Also, machine learning potentials often struggle to accurately capture substantial bond stretching events \cite{f.dossantosImprovingBondDissociations2025}. 
Accordingly, we introduce a new deep learning-based PES for molecular dynamics simulations of solid CO, including its vibrational excitation in surfaces and aggregates. This PES is rigorously trained on DFT data acquired from our previous studies \cite{delfreMechanismUltravioletInducedCO2023}. In contrast to the conventional approach of employing a single neural network (NN) 
for each chemical element, we propose an alternative strategy in which atoms of the vibrationally excited CO molecule are modeled using separate networks 
to address the possible limitations mentioned above in order to enhance stability. The following discussion outlines the construction and validation of the MLP for CO using DeePMD-kit~\cite{WangDeePMDkitdeeplearning2018, ZengDeePMDkitv2software2023}, followed by its application in large-scale molecular dynamics simulations to investigate vibrational relaxation in the context of photodesorption in CO ice analogues. Internal and translational energy distributions of the desorbed species, obtained from 11,000 trajectories, will be analyzed and compared with experimental results as well as with those previously derived from AIMD simulations.

\section{Methods}

\subsection{Neural network potential energy surface}

The MLP was constructed using the open-source DeePMD-kit package~\cite{WangDeePMDkitdeeplearning2018, ZengDeePMDkitv2software2023} 
following the Behler-Parrinello approach~\cite{PhysRevLett_98_146401}, in which the total energy of the system $E_{\text{tot}}$ is obtained by summing over the energies of all individual atoms, i.e., $E_{\text{tot}} = \sum_i E_i$, where $E_i$ represents the atomic energy contribution of the $i$th atom.
In this framework, the MLP is composed of a descriptor neural network and fitting neural networks.  
The MLP descriptor network was built using the two-body embedding DeepPot-SE descriptor~\cite{ZengDeePMDkitv2software2023,WenDeeppotentialsmaterials2022} .
It consists of three layers, each containing 50 nodes associated to a hyperbolic tangent as activation function, the cut-off radius was set to 6.5~\AA{}, and the size of the submatrix is set to 8.

The fitting neural network also consists of three hidden layers, each containing 250 neurons and similarly employing a hyperbolic tangent activation function. During training, the learning rate smoothly decays from $10^{-2}$ to $10^{-3}$. The weights applied to the energy and force loss functions evolve dynamically throughout training, beginning with initial values of 1 and 1000, respectively, and progressively adjusting to final values of 8 and 1.
Training was conducted over 400,000 steps using the ADAM optimization algorithm~\cite{kingma2014adam}, with a batch size of 1. Five distinct MLPs were trained with identical hyperparameters but different initial neural network weights. The MLP demonstrating the lowest training and validation errors was selected for subsequent molecular dynamics (MD) simulations.

The dataset employed for training and validating the MLP was derived and expanded from previous research~\cite{delfreMechanismUltravioletInducedCO2023}. All calculations were performed using the density functional theory (DFT)-based Vienna Ab Initio Simulation Package (VASP)~\cite{Kresse_96_1, Kresse_96_2} (version 6.2 and 6.3), employing the Perdew-Burke-Ernzerhof (PBE) exchange-correlation functional~\cite{Perdew_96} with the DFT-D3(BJ) dispersion correction~\cite{Grimme_2010,Grimme_2011} to adequately describe van der Waals interactions. Ionic cores were described using the projector augmented wave (PAW) method~\cite{bl1994p} implemented in VASP~\cite{Kresseultrasoftpseudopotentialsprojector1999a}.
Aggregates consisting of 50 CO molecules were generated by randomly positioning molecules inside an initial sphere of radius 8.2~\AA{} with random orientations. After structural optimization, aggregates had an average radius of approximately 8.6~\AA{}. The ensemble of aggregate configurations obtained through canonical ensemble thermalization, which recorded positions and velocities of CO molecules, was modified by introducing vibrational excitation to a single randomly selected CO molecule situated within a central radius of approximately 4.5~\AA{}. The internal momentum of this molecule was set to match the selected vibrational quantum number $v = 40$ in this work.
For each AIMD trajectory, an aggregate configuration featuring an excited CO molecule was randomly selected from the available set. Constant-energy AIMD simulations were conducted using the Verlet algorithm implemented in VASP, resulting in a total of 100 trajectories, among which desorption was observed in 88~\% of the cases.

For the whole study, we have computed more than 700 000 configurations. Thus, a sampling strategy has to be done in order to reduce the amount of configuration to start training our model. The process of selection is composed of three steps for each trajectory. First, from the beginning of the AIMD until $t =$ 500~fs, we select one configuration every 100~fs. Then, between 100~fs and 2500~fs, we select a configuration every 10~fs. The primary motivation for increasing the selection frequency in this region is that it captures the key physical phenomena, ensuring an accurate description by our model. Finally, for configurations above 2500~fs, we select a configuration every 100~fs. Overall, the resulting dataset is composed by a total of 18 000 configurations.

To overcome the challenges inherent to the modeling of vibrationally excited molecules, we have developed an innovative strategy in which the vibrationally excited CO molecule is uniquely described by distinct neural network entities labeled X$_1$ and X$_2$. This novel approach allows us to explicitly differentiate the excited molecule from the surrounding ground-state molecules, thus significantly enhancing the MLP's stability and accuracy. While conventional neural network potentials treat all atoms of the same element indistinctly, our method explicitly distinguishes the atoms of the excited CO molecule. By doing so, it robustly handles highly anharmonic and energetic conditions specific to vibrationally excited states, marking a notable methodological advancement in MLP construction.
As mentionned before, we have used a atomic cut-off radius $r_{c}$ of 6~\AA{}, associated to a smooth cut-off radius of 0.5~\AA{}, leading to a total cut-off radius of 6.5~\AA{}. For carbon and oxygen atoms, we have set the maximum of neighbor atoms to 29. Regarding X$_1$ and X$_2$ atoms, we have set this number to 1, knowing that they are unique in the system. For the training procedure, the whole data set is distributed randomly with 80 \% of the configurations for the training set and 20 \% for the test set. An additional validation set composed of 2000 randomly selected configurations is used to validate the MLP predictions. 

\subsection{Molecular dynamics}

Molecular dynamics (MD) simulations were carried out using the Large-scale Atomic/Molecular Massively Parallel Simulator (LAMMPS) code~\cite{plimpton_2023_10806852} associated to its module to incorporate MLP models. The MLPs were trained to increase the statistics we made in our previous AIMD study~\cite{delfreMechanismUltravioletInducedCO2023}. Thus, we applied the exact same methodology. 

First, we optimize a 50 CO aggregate to obtain the initial configurations. The initial configurations containing only ground state CO molecules, neither X$_1$ or X$_2$ atoms were used until the production step. In order to be representative of a canonical ensemble at 15~K, we thermalized the system using a Nosé-Hoover~\cite{Noseunifiedformulationconstant1984,HooverCanonicaldynamicsEquilibrium1985} thermostat, setting its coupling constant to 10 fs. The equations of motions were integrated using the velocity form of the Stoermer-Verlet time integration algorithm~\cite{VerletComputerExperiment1967,SwopeComputerSimulationMethod1982} with a time step of 0.1 fs over 5 ns.

\begin{figure}[h!]
    \centering
    \includegraphics[width=0.8\linewidth]{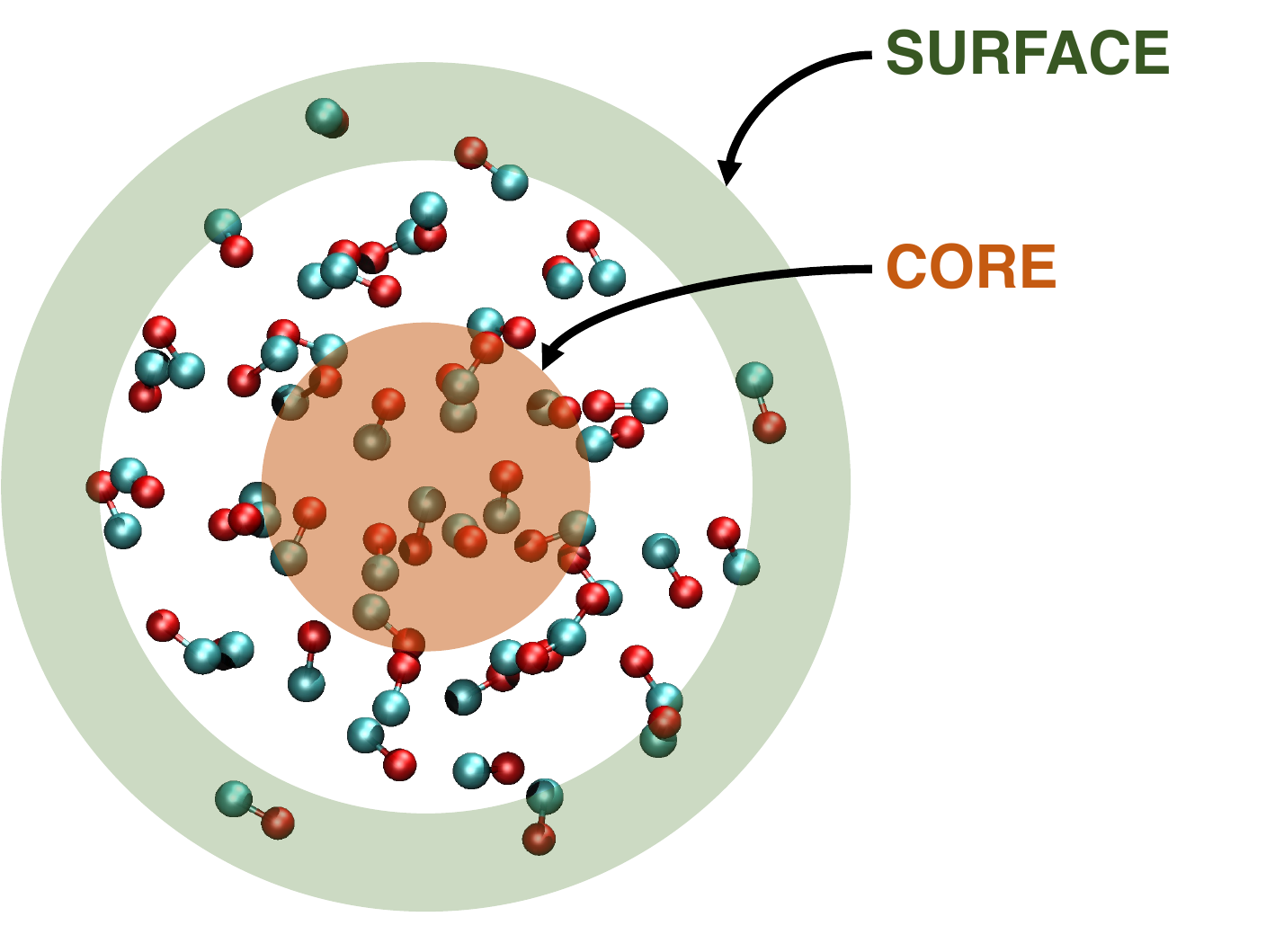}
    \caption{Representation of one CO aggregate. The core region, represented in red, is a sphere centered on the aggregate center of mass with a radius of 4.5~\AA{}. The surface region, represented in green, represents the molecules at the interface.}
    \label{fig:excitation_scheme}
\end{figure}

The configurations obtained during the later phase will serve as initial conditions for the MD calculations. Indeed, we select randomly 11 000 configurations from the previous simulation having their instantaneous temperatures ranging between 13 and 17 K. For each configuration, one single CO molecule is picked to be vibrationally excited inside the aggregate, i.e., having its center of mass at a maximum distance of 4.5~\AA{} from the aggregate center of mass as shown in Fig.~\ref{fig:excitation_scheme}.

This molecule will have its internal momentum modified to match a ground vibrational state corresponding to $v = 40$. The carbon and oxygen atoms from the excited molecule become respectively X$_1$ and X$_2$ due to our MLP model design.
With 11 000 different initial conditions set, we run MD simulations in the micro-canonical ensemble using the LAMMPS code and the MLP model to compute energies and forces. The simulations run over 5 ps with a time-step set to 0.1~fs using the same algorithm than the thermalization step to integrate the equations of motion. Two possible exit channels were taken into account, called desorption and non-desorption: a molecule was considered desorbed when the distance from the CO center of mass and the aggregate surface exceeded 3~\AA{} and not-desorbed when after 5~ps no molecule fulfilled the above condition. The total energy was well conserved for each trajectory with a standard deviation of $\sim$~0.98~meV. The energy distributions of the molecules along the dynamics were obtained using the standard semi-classical determination method for the translational, vibrational and rotational energies \cite{billing2000BooK}.

\section{Results}

\subsection{Neural network validation}

\begin{figure}[h]
    \centering
    \includegraphics[width=0.8\linewidth]{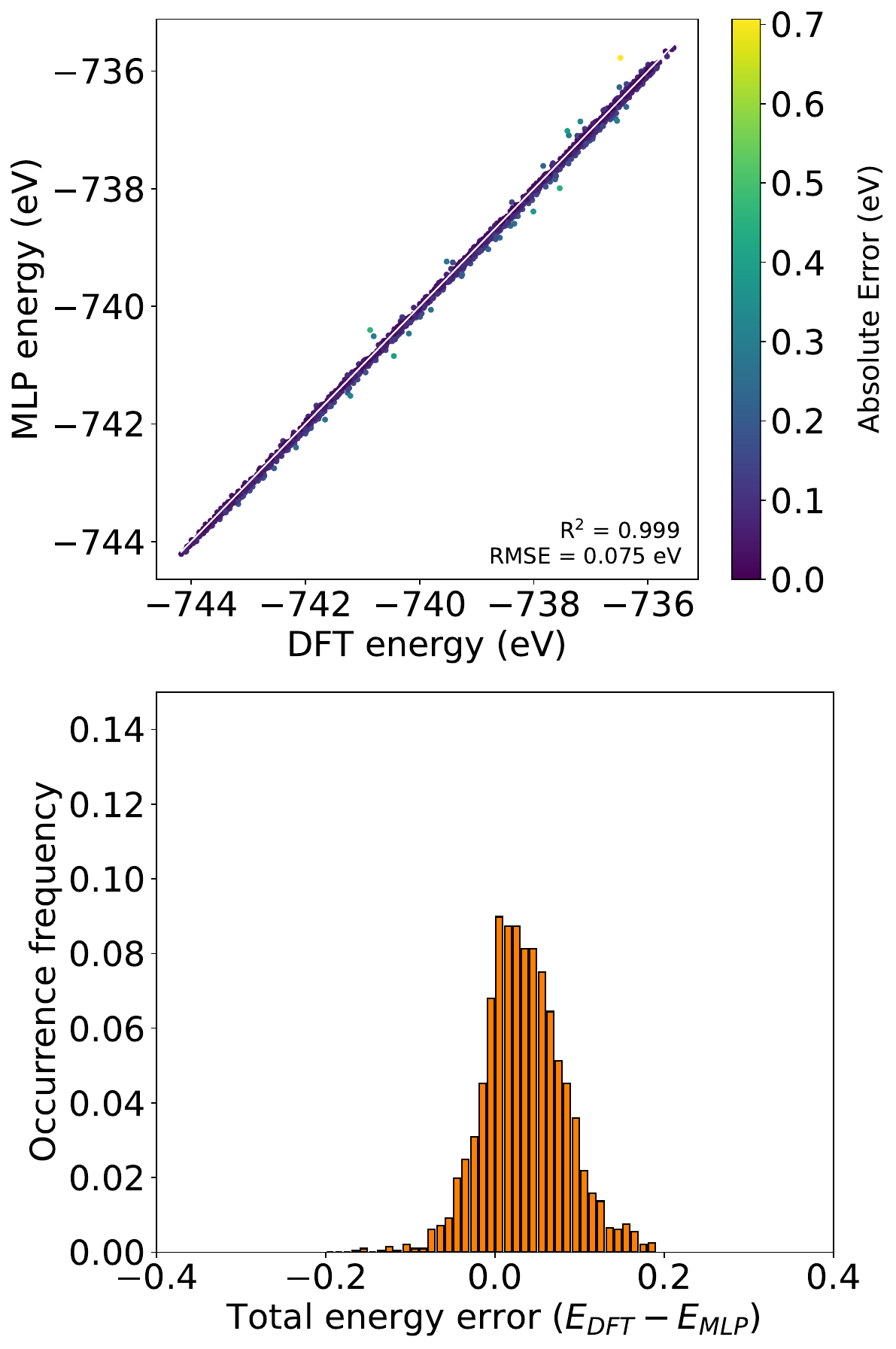}
    \caption{Parity plot of the total energy values (top panel) and the distribution of errors (bottom
    panel) obtained by the MLP prediction of the DFT set. The bin size used in the
    errors histogram is 0.01 eV. The total root mean square error on the energies is 75~meV.}
    \label{fig:parity_energy}
\end{figure}

When building a neural network PES, ones necessarily needs to check the accuracy of predictions of the energies and the forces on your dataset. Regarding the energy predictions during the training, we end up with a root mean square error (RMSE) of 0.5~meV/atom on the training set and 0.1~meV/atom for the testing set. Then, we have compared the energies of the AIMD calculations with the ones computed using our MLP model of the validation set in Fig.~\ref{fig:parity_energy}. We observe a good correlation between the energies predicted by the neural network and DFT, where the RMSE of our MLP is 75~meV. In addition, having 100 atoms in our system, it lowers the atom normalized RMSE at 0.8~meV. Such errors are comparable to other published MLP based on the same architecture~\cite{BehlerGeneralizedNeuralNetworkRepresentation2007,BehlerAtomcenteredsymmetryfunctions2011,HedmanDynamicsgrowingcarbon2024,ZhangPolarizationdrivenbandtopology2024, ToselloGardiniMachinelearningdrivenmolecular2025}. Moreover, the standard deviation on the errors is 60~meV, with most of the errors ranging between -0.1 and 0.2~eV for the whole system. It corresponds to a relative error of 0.04~\% of the system energy and 3.63~\% of the initial vibrational energy of the excited CO molecule. Concerning the errors distribution in the bottom panel, we observe a subtly asymmetry of the distribution. It indicates a tendency from our MLP to slightly overestimate DFT energies. Nevertheless, such differences with DFT energies would have minor impact on molecular dynamic simulations using our MLP model.

\begin{figure*}[ht]
    \centering
    \includegraphics[width=\linewidth]{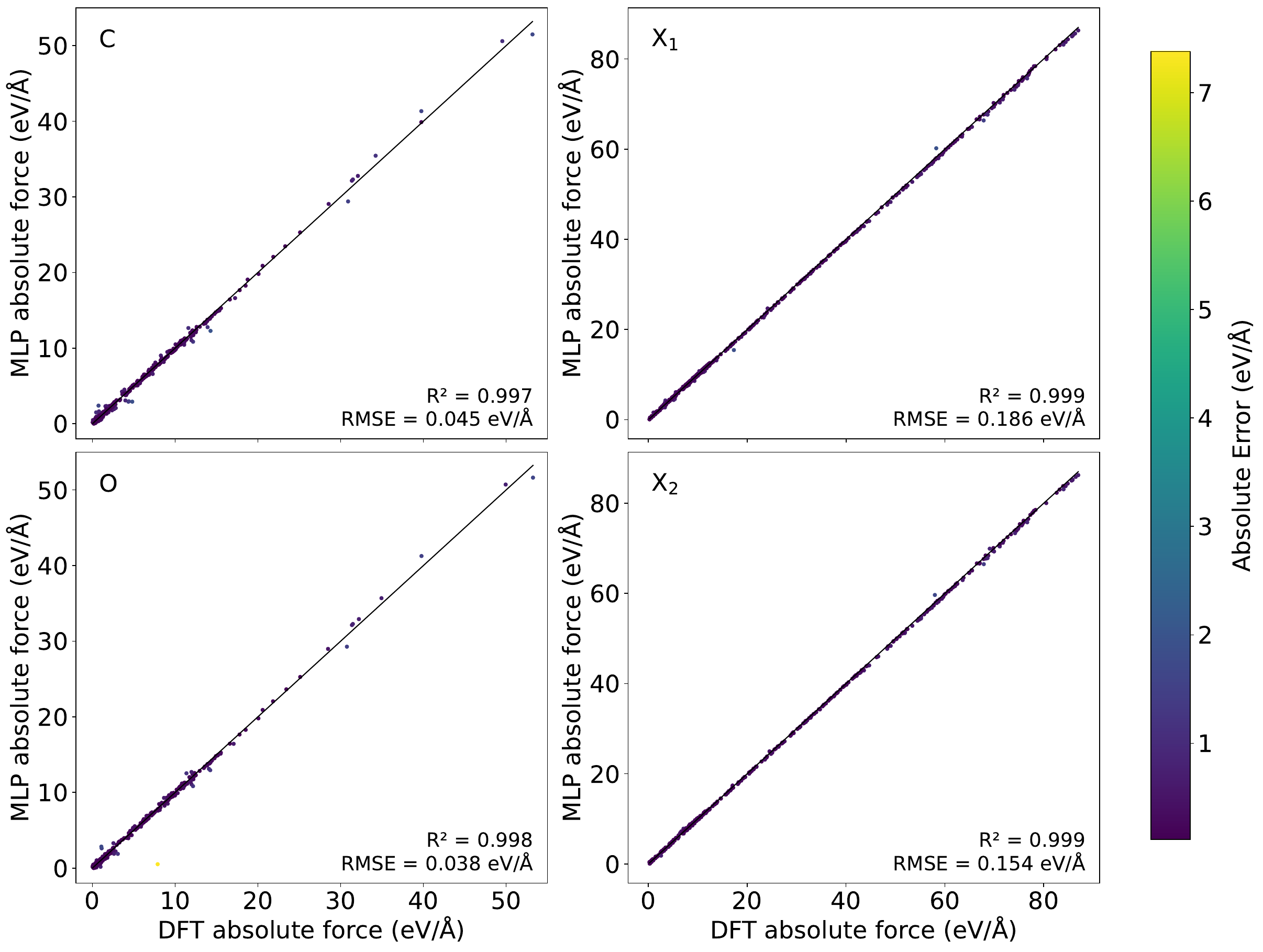}
    \caption{Parity plot of the absolute value of the forces obtained by the MLP prediction of the DFT set for each X$_1$ (top right pannel, excited carbon atom), X$_2$ (bottom right pannel, excited oxygen atom), C (top left pannel, remaining carbon atoms),  and O (bottom left pannel, remaining oxygen atoms) atoms involved. The colorbar shows the absolute error between MLP and DFT forces.}
    \label{fig:parity_force}
\end{figure*}

\begin{figure*}[ht]
    \centering
    \includegraphics[width=\linewidth]{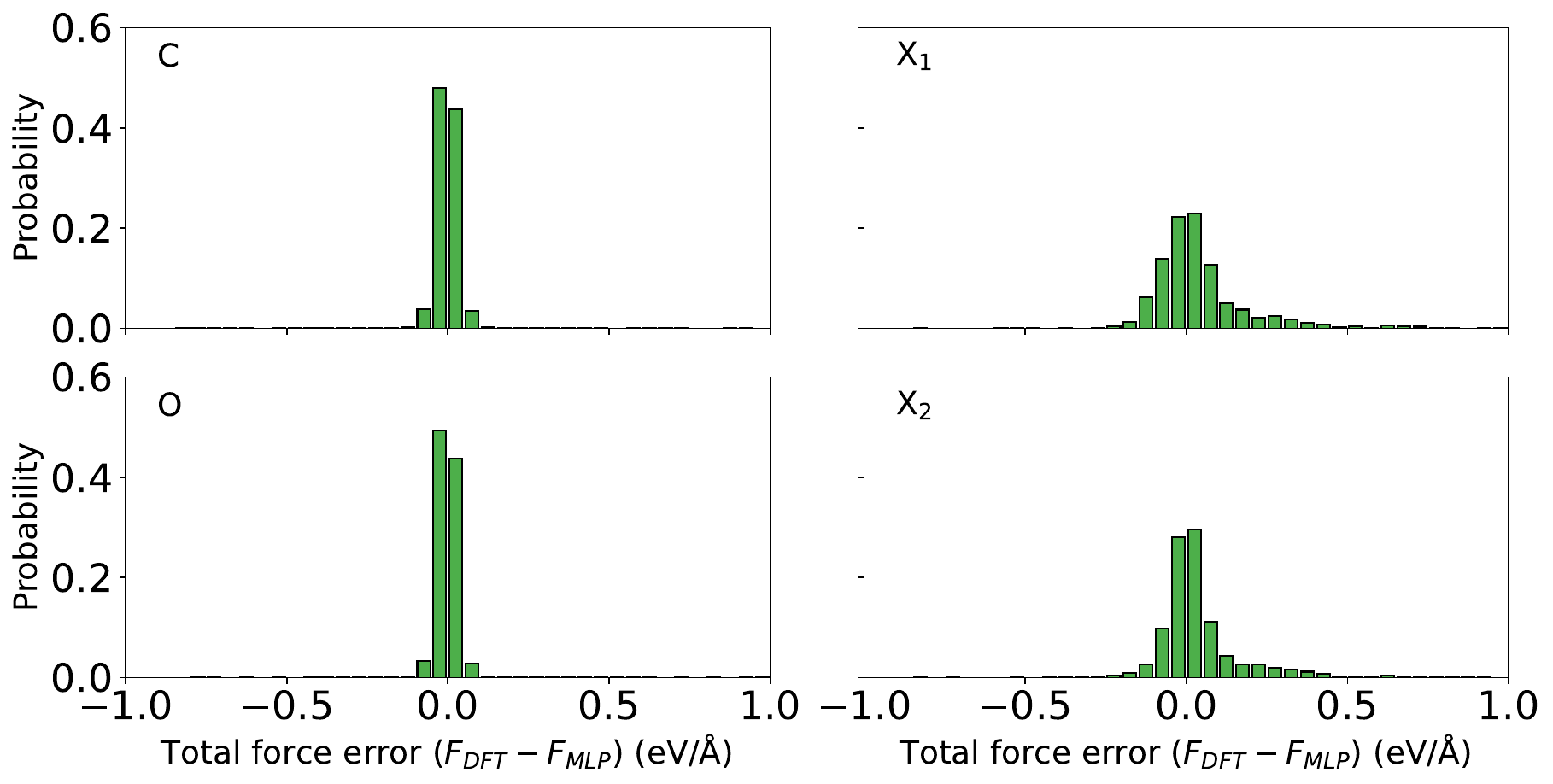}
    \caption{Distribution of the absolute error of the forces obtained by the MLP prediction of the DFT set for each atom involved. The bin size used in the histogram is 0.05 eV/\AA.}
    \label{fig:parity_force_distribution}
\end{figure*}

Moving on the force predictions, we show the same comparison as for the energies in Fig.~\ref{fig:parity_force} and Fig.~\ref{fig:parity_force_distribution} for each atom involved in the MLP. Regarding the training, we end up with a overall RMSE of 0.038 eV/\AA{}. For the validation set, we observe an excellent correlation between carbon and oxygen atom forces predicted by the neural network and DFT, with a RMSE of 0.045~eV/\AA{} and 0.038~eV/\AA{} respectively. Regarding X$_1$ and X$_2$ results, we observe a good correlation between forces predicted by our MLP model and DFT, with a RMSE of 0.186~eV/\AA{} and 0.154~eV/\AA{} respectively. Even if the RMSE error is larger than the one for carbon and oxygen atoms, it is still an excellent result regarding the absolute force range that is predicted. Indeed, most of the absolute forces for carbon and oxygen atoms are below 40~eV/\AA{} while X$_1$ and X$_2$ absolute forces range between 0 and 80~eV/\AA{} homogeneously. Therefore, a small relative error of the force can be interpreted as a high absolute error. In addition, for each configuration, we have 1 X$_1$ and X$_2$ atom for 49 carbon and oxygen atoms. Regarding the distribution of the absolute error in Fig.~\ref{fig:parity_force_distribution}, we confirm an excellent correlation for force prediction on each atom. The same asymmetry discussed for the energy error distribution is found for X$_1$ and X$_2$ atoms, with a tendency of underestimating DFT forces slightly.

\begin{figure*}[ht]
    \centering
    \includegraphics[width=\linewidth]{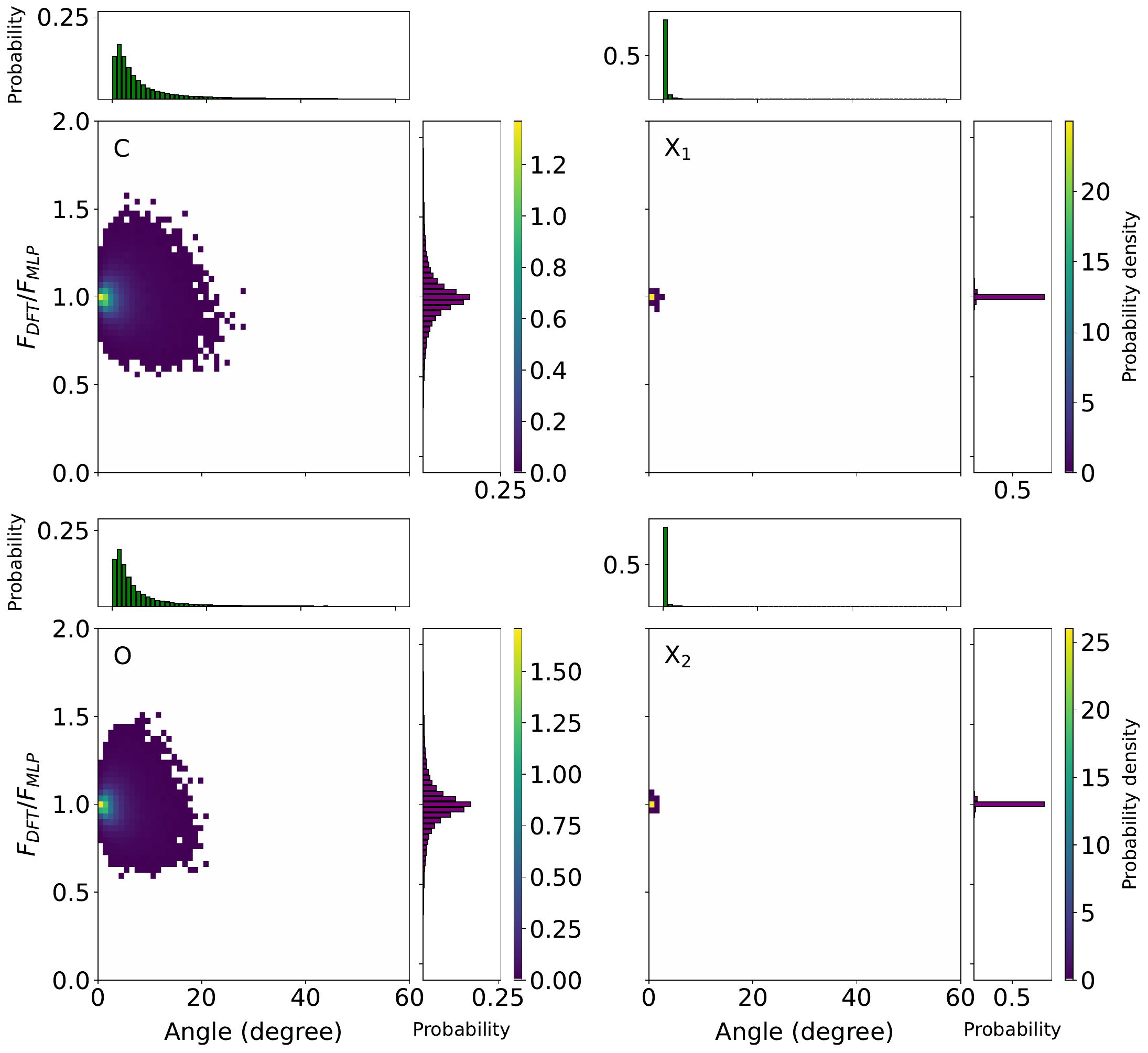}
    \caption{Heatmap representing the relative force error and the angle between DFT and MLP predicted force vector. The highest density is colored in yellow and the lowest one in dark purple. The distribution of the relative error (purple) is shown on the right, and the distribution of angles (green) on the top.}
    \label{fig:heatmap_force}
\end{figure*}

To improve the analysis of the accuracy of our MLP model force predictions, we not only need to compare the absolute error on them, but also the direction of the force vector as well as the relative error made on them. For this purpose, we have compared the angle between DFT and MLP force vectors for each atoms, as well as the relative error made by the MLP model on a heatmap shown in Fig.~\ref{fig:heatmap_force}. We observe an excellent correlation for carbon and oxygen atoms, with a majority of angle difference below 5$\degree$ and a low relative error on predicted forces. Moving on X$_1$ and X$_2$ atoms, we obtained a perfect correlation between MLP predicted forces and DFT ones. Indeed, all of the points are located at the same position on the heatmap. This result can be explained by the behavior of X$_1$ and X$_2$ atoms, having high forces value along the same relative direction along all DFT calculations. Nevertheless, this analysis confirms that the absolute error distribution obtained in Fig.~\ref{fig:parity_force_distribution} as well as its RMSE are not an indicator of a low accuracy on the X$_1$ and X$_2$ force predictions. 

\begin{figure}[ht]
    \centering
    \includegraphics[width=\linewidth]{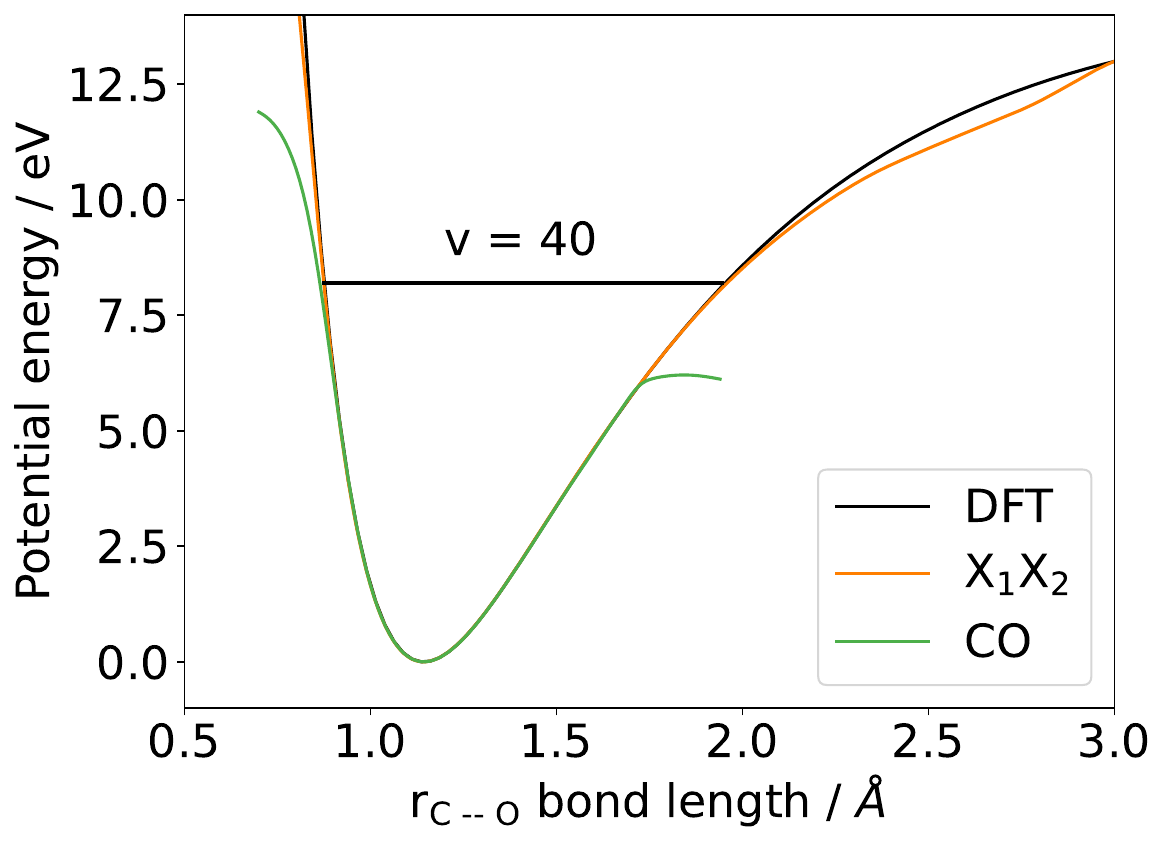}
    \caption{Comparison of the dimer dissociation computed at the DFT non spin-polarized level of theory (black) with X$_1$X$_2$ (orange) and CO (green) MLP predictions. The potential energy corresponding at $v = 40$ (corresponding energy E~=~8.26~eV) is indicated by a horizontal black line.}
    \label{fig:dissociation_curve}
\end{figure}

Finally, the last check we need to perform on the MLP is to check if it reproduces well the DFT dissociation curve. Results are shown in Fig.~\ref{fig:dissociation_curve}. Note that the training set contains only values below 8.26~eV for X$_1$X$_2$, as it corresponds to the initial condition of the excited CO molecule. In addition, MLP's carbon and oxygen atoms only describe ground state CO molecules, limiting even more the region in which they have information on the CO dissociation curve. We observe an excellent corelation between the DFT dissociation curve and X$_1$X$_2$ for potential energies below 8.26~eV. Moreover, predictions made by CO atoms are also in excellent agreement with the DFT dissociation curve for values below 6.5~eV. This result can be explained by the vibrational excitation of CO molecules induced by the kick from X$_1$X$_2$ during the dynamics.

Overall, the excellent results obtained when assessing the accuracy of our MLP predictions indicates that the later is sufficiently accurate to predict energies and forces, enabling us to perform MD while lowering the computational cost and keeping a good accuracy. The introduction of X$_1$ and X$_2$ to represent the excited molecule significantly improved the accuracy of the MLP predictions, despite the inherently challenging vibrational excitation scenario. This demonstrates the robustness of our novel approach in accurately capturing extreme vibrational states. In other words, we are able to increase significantly the number of MD trajectories while keeping an excellent accuracy on energies and forces, leading to a significant increase of the statistical results obtained from our previous study~\cite{delfreMechanismUltravioletInducedCO2023}.

\subsection{Molecular Dynamics Simulations}

Using the validated MLP, we ran 11\,000 trajectories using LAMMPS under the above-described initial conditions, i.e. 1100 times the number of trajectories performed in our previous AIMD calculations using VASP. Among these 11\,000 trajectories, 77~\% resulted in the desorption of CO molecule(s). This result is close to that predicted by our previous AIMD calculations \cite{delfreMechanismUltravioletInducedCO2023} (88~\%) , demonstrating that the new MLP provides an accurate and reliable representation of the system’s multidimensional potential energy landscape. The average desorption time was 2~ps, and no desorption was observed before 600~fs, confirming the previous AIMD and experimental findings but now with improved statistical sampling. The new simulations confirm that when a molecule at the center of the aggregate is excited, no immediate release into the gas phase occurs. Instead, the desorption mechanism proceeds indirectly through three main steps. First (the vibration), the excited molecule vibrates inside the aggregate while retaining the initially deposited vibrational energy. Second (the kick), the excited molecule and one or two CO molecule(s) in its vicinity begin to be mutually attracted, leading to their collision. Third (the desorption), the colliding molecules start moving and interacting with other molecules within the aggregate, leading to a cascade energy transfer effect. This process ultimately provides the surface CO molecule(s) with sufficient kinetic energy to exceed the binding energy of the aggregate.

\subsection{Desorbed molecules energy distributions}

Building on the extensive dataset generated by MD simulations employing the validated MLP, a comprehensive analysis of the energy distributions of the desorbed molecules is presented. The substantially enhanced statistical sampling, relative to previous AIMD studies, affords improved accuracy and provides deeper insight into the desorption dynamics.

\begin{figure}[ht]
    \centering
    \includegraphics[width=\linewidth]{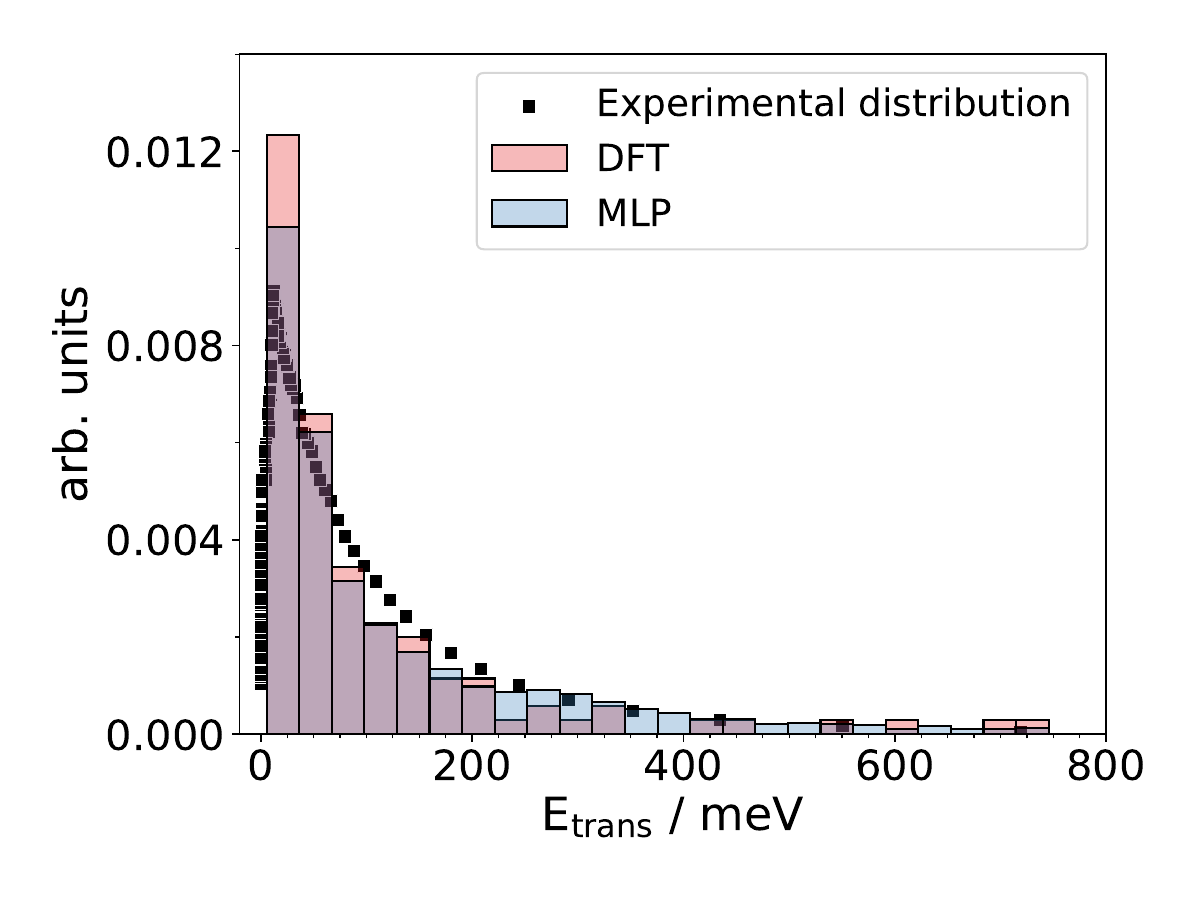}
    \caption{Translational energy distributions computed using MLP (blue) compared with previous DFT (red) and experimental data (black) from Del Fré \textit{et. al.} \cite{delfreMechanismUltravioletInducedCO2023} The purple color of the histograms corresponds to the superposition of the blue and the red histograms. If the purple histogram is surmounted by the red one, then it corresponds to MLP calculations, otherwise it corresponds to DFT calculations.}
    \label{fig:E_trans_exp}
\end{figure}

As previously observed in both experimental and theoretical studies \cite{delfreMechanismUltravioletInducedCO2023}, the MLP MD simulations show that the photodesorbed CO molecules are predominantly in their ground vibrational level ($v=0$), with the population of the $v=1$ level estimated at approximately 5\%. Figure~\ref{fig:E_trans_exp} shows the experimental and theoretical translational energy distributions (AIMD and MLP) of the desorbed molecules in the $v=0$ and $J<7$ states.

The kinetic energy distributions show a peak centered around 25~meV, with very few molecules reaching translational energies above 200~meV. Notably, the distributions obtained from the MLP simulations are significantly smoother than those derived from AIMD, owing to the much larger number of trajectories. Furthermore, the MLP results accurately reproduce the entire range of translational energies detected experimentally, including the pronounced decrease in the probability of detecting highly translationally excited molecules. In agreement with the experimental observations, the new theoretical results confirm that no molecules with translational energies above 800~meV were detected. This improved statistical resolution enables a more reliable and detailed comparison with the experiment and further supports the accuracy of the MLP in capturing the desorption dynamics. The excellent agreement observed across both vibrational and translational energy distributions underscores the robustness of the description of the photodesorption mechanism initially proposed experimentally and subsequently confirmed by our simulations \cite{delfreMechanismUltravioletInducedCO2023, hacquardPhotodesorptionCOIces2024}.

When focusing on the rotational energy distributions, the advantage of the improved statistics afforded by the MLP simulations becomes even more evident. Experimentally, the rotational distributions were extracted within specific windows of translational energy, which made direct comparison with AIMD results challenging. Due to the limited number of AIMD trajectories, broader translational energy ranges had to be used in order to achieve sufficient sampling \cite{hacquardPhotodesorptionCOIces2024}. In contrast, the large dataset generated with the MLP allows for a much finer selection of desorbed molecules according to their translational energy, enabling a more accurate and meaningful comparison with the experimental distributions.

\begin{figure}[ht]
    \centering
    \includegraphics[width=0.8\linewidth]{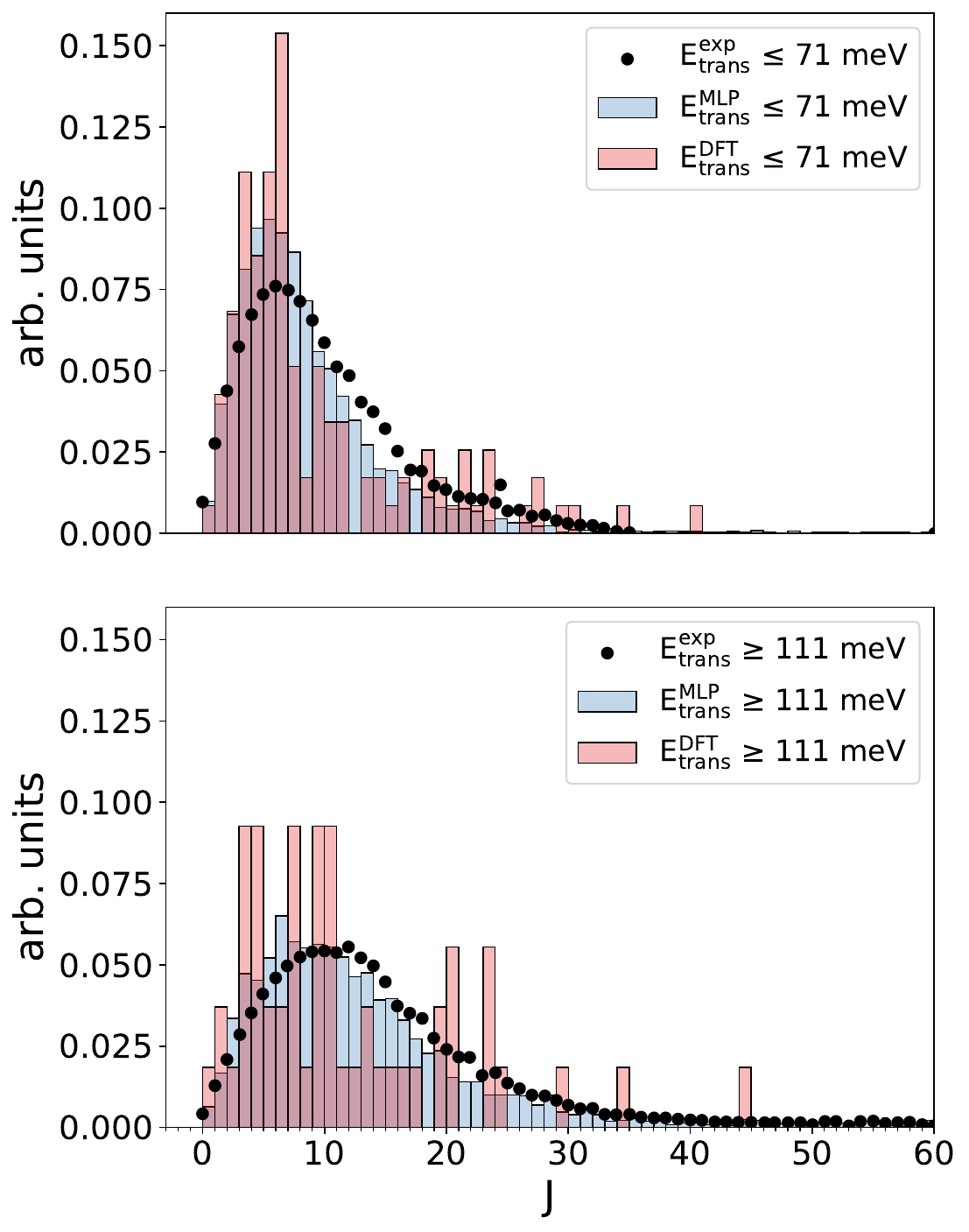}
    \caption{Rotational energy distributions at low (top panel) and high translational energy values (bottom panel) among all the trajectories using MLP (blue) and DFT (red) theoretical methods compared to experimental data from Hacquard \textit{et. al.}~\cite{hacquardPhotodesorptionCOIces2024} (black). High values distributions are computed for translational energies lower than 200 meV. All the distributions are density probabilities.}
    \label{fig:J_etrans_combined}
\end{figure}

Figure~\ref{fig:J_etrans_combined} presents a comparison between the experimental data and the theoretical results from both AIMD and MLP simulations for two translational energy intervals: one corresponding to molecules with low translational energy (below 71~meV) pannel (a), and another with higher translational energy (above 111~meV) pannel (b). While these ranges are similar to those used in our earlier work \cite{hacquardPhotodesorptionCOIces2024}, the improved statistics afforded by the MLP now allow us to match the experimental energy intervals precisely, without the need to artificially broaden them to obtain sufficient sampling. The comparison clearly shows that the MLP simulations markedly improve the theoretical description of the rotational distributions. In both energy ranges, the MLP results closely follow the experimental trends, providing smoother distributions with reduced statistical noise. As observed for the translational energy distribution, the MLP-based rotational distributions also capture the characteristic decay of the experimental profiles across both energy intervals with high fidelity. A closer inspection of the rotational distributions reveals that, for desorbed molecules with translational energies below 71~meV, the distribution peaks around $J \sim 5$, and the probability of finding rotational states with $J > 35$ is very low. Similarly, for molecules with translational energies above 71~meV, the rotational distribution exhibits a maximum around $J \sim 10$,  with the population of states with $J > 40$ becoming negligible.

\begin{figure*}[ht]
    \centering
    \includegraphics[width=\linewidth]{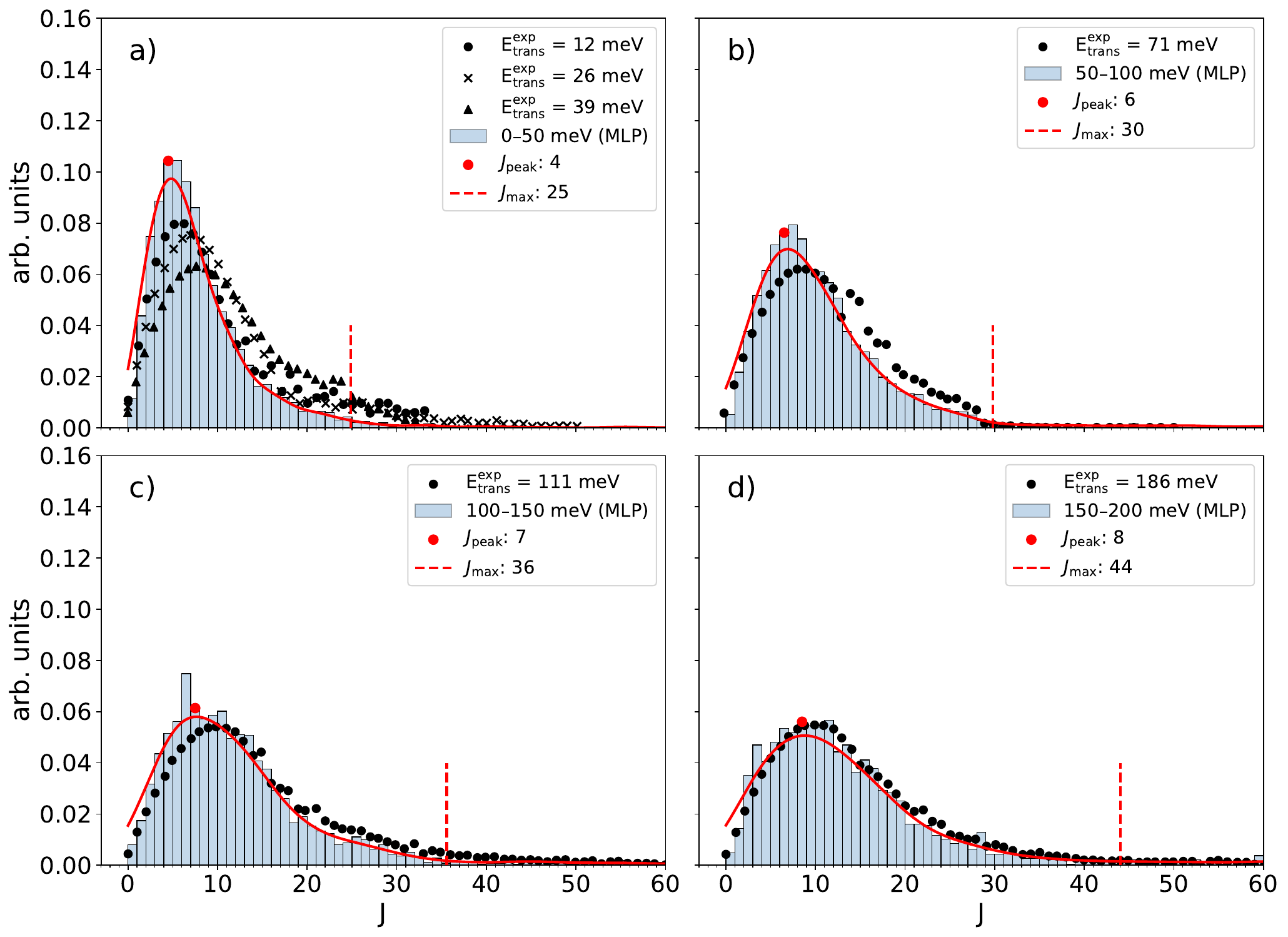}
   
\caption{Theoretical rotational MLP energy distributions (blue) compared to experimental data~\cite{hacquardPhotodesorptionCOIces2024}(black) for four translational energy ranges. Panel a : [0 mev 50 mev],  Panel b : [50 mev 100 mev],  Panel c : [100 mev 150 mev], and  Panel d : [150 mev 200 mev]. To estimate the position of the distribution maximum $J_\text{peak}$ (red dot), we used the kernel density estimation (KDE) represented as a red line. The end of each distribution (represented by a red dotted line) $J_\text{max}$ is empirically defined when the density falls below 3\% of the distribution maximum.}    
    \label{fig:jdist_vs_etrans}
\end{figure*}
    
Another advantage of the improved statistical sampling provided by the MLP MD simulations is the ability to study rotational distributions over narrower translational energy intervals. Figure~\ref{fig:jdist_vs_etrans} shows the rotational distributions for specific translational energy ranges: 0–50~meV (panel (a)), 50–100~meV (panel (b)), 100–150~meV (panel (c)), and 150–200~meV (panel (d)). The theoretical rotational distributions are fitted using kernel density estimation (KDE)~\cite{2020SciPy-NMeth} in order to determine $J_\text{peak}$ and $J_\text{max}$, corresponding respectively to the rotational quantum number associated with the position of the distribution maximum and the beginning of the distribution tail, which we defined as the position at which the KDE fit reaches 3\% of its maximum value. Notably, such rotational distributions could not be reliably obtained from AIMD simulations due to insufficient statistics within each energy interval to ensure convergence. This highlights a major advantage of the MLP MD simulations over AIMD. The first important outcome is that the theoretical distributions are smooth, indicating that the sampling is statistically robust. The second is the excellent agreement observed between theory and experiment.

\begin{figure}[ht]
    \centering
    \includegraphics[width=0.8\linewidth]{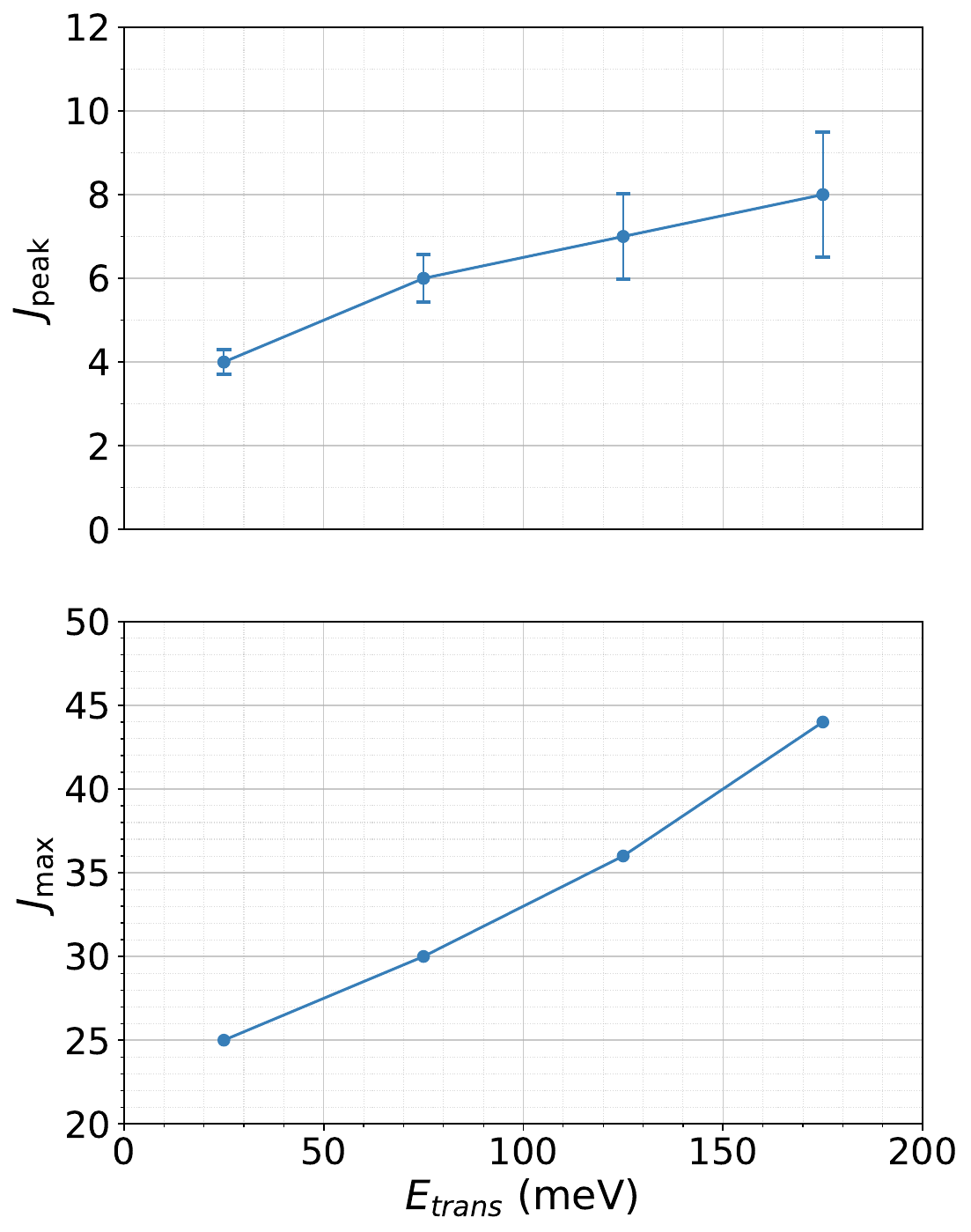}
    \caption{Evolution of the most probable $J_{\mathrm{peak}}$ (top pannel) and the estimated highest populated $J_{\mathrm{max}}$ (bottom pannel) rotational quantum number according to translational energy. The points correspond to the same intervals shown in Figure~\ref{fig:jdist_vs_etrans}. Bars in the top panel correspond to the uncertainty on $J_{\mathrm{peak}}$ within a 95~\% interval.}
    \label{fig:jdist_vs_etrans2}
\end{figure}

The improved statistical sampling resulting from the MLP MD simulations enables a more detailed investigation of the correlation between translational and rotational energies, a remarkable characteristic of the CO photodesorption mechanism previously discussed in our earlier work \cite{hacquardPhotodesorptionCOIces2024}, where both experimental and AIMD results indicated that translational and rotational energy quantities vary jointly. The Figure~\ref{fig:jdist_vs_etrans2} shows the dependence of $J_\text{peak}$ and $J_\text{max}$ as a function of the mean translational energy for each selected interval. 

Their evolution provides evidence of a correlation between rotational and translational degrees of freedom during the CO desorption process. Both the most probable and the highest populated rotational quantum numbers increase regularly with translational energy, following a nearly linear trend across the full energy range studied. This behavior indicates a continuous and well-coupled energy transfer mechanism, where rotational excitation scales proportionally with the kinetic energy acquired by the desorbing molecule. Such sustained growth highlights the efficiency of energy redistribution and supports the picture of a dynamically coupled system where rotational and translational motions evolve coherently throughout the desorption event.

\section{Conclusion}

In summary, we have developed a high-dimensional machine learning potential capable of accurately describing the CO desorption process following the vibrational excitation ($v=40$) of a single CO molecule within an aggregate of 50 CO molecules. The MLP was validated by comparing its predictions with a large reference dataset obtained from DFT calculations. The results showed low errors in both energies and forces predicted by the MLP, as well as excellent agreement with our previous AIMD study. The molecular dynamics simulations performed with LAMMPS are approximately $\sim$~1500 times faster in computing a single point (energy and forces) compared to our previous AIMD calculations.

An important aspect of the present work is the successful modeling of the vibrationally excited CO molecule using two separate neural networks, one for each atom (X$_1$ and X$_2$). This architecture allowed the MLP to accurately capture the strong anharmonicity and asymmetry associated with the high vibrational excitation ($v=40$), while maintaining a consistent description of the surrounding non-excited molecules. The ability to selectively treat the excited molecule within a heterogeneous condensed-phase environment represents a significant methodological advance and opens the door to modeling other localized excitations in complex molecular systems.

The MLP MD simulations accurately reproduced the high desorption yield of CO molecules, in excellent agreement with previous AIMD predictions. This result aligns with experimental observations, since in both cases, the large majority of UV excitations (in experiments) or vibrational excitations (in simulations) of CO molecules result in a desorption event. The significantly improved statistical sampling enabled by the MLP also provided a much finer description of the translational energy distributions, allowing a more detailed comparison with experimental time-of-flight spectra. Furthermore, the analysis of the rotational energy distributions, resolved over specific translational energy intervals, revealed a clear linear correlation between translational and rotational excitation. In particular, both $J_\text{peak}$ and $J_\text{max}$ were found to increase proportionally with translational energy across the studied range, highlighting the efficiency of energy redistribution and the strongly coupled nature of the desorption dynamics. This finding provides a deeper understanding of energy partitioning mechanisms during vibrationally induced desorption from molecular ices.

Building on these achievements, the MLP developed here offers a robust and highly efficient framework for future studies of photodesorption dynamics. The combination of high-dimensional neural network potentials with selective treatment of the excited molecule has proven to be a powerful strategy for capturing complex energy redistribution dynamics in condensed-phase environments. The remarkable computational efficiency of the MLP opens new perspectives for extensive statistical studies on desorption energies, detailed investigations of surface molecule excitations, and the exploration of larger-scale models incorporating periodic boundary conditions to simulate more realistic ice surfaces. Future work based on this potential will thus allow a deeper and more comprehensive understanding of vibrationally induced desorption mechanisms under conditions closer to those found in astrophysical and surface science contexts.

\begin{acknowledgments}

This work was funded by the ANR PIXyES project, grant ANR-20-CE30-0018 of the French ‘‘Agence Nationale de la Recherche’’ and ANR SFRI (21-SFRI-0005) GRAEL program of the University of Lille. 
The authors acknowledge support from the French national supercomputing facilities (Grant Nos. DARI A0130801859, A0110801859) and from the Centre de Ressources Informatiques (CRI) of the Université de Lille. 
The French State under the France-2030 programme and the Initiative of Excellence of the University of Lille are acknowledged for the funding and support granted to the R-CDP-24-003-AREA project.
\end{acknowledgments}

\bibliography{biblio_JCP}

\begin{thebibliography}{56}%
\makeatletter
\providecommand \@ifxundefined [1]{%
 \@ifx{#1\undefined}
}%
\providecommand \@ifnum [1]{%
 \ifnum #1\expandafter \@firstoftwo
 \else \expandafter \@secondoftwo
 \fi
}%
\providecommand \@ifx [1]{%
 \ifx #1\expandafter \@firstoftwo
 \else \expandafter \@secondoftwo
 \fi
}%
\providecommand \natexlab [1]{#1}%
\providecommand \enquote  [1]{``#1''}%
\providecommand \bibnamefont  [1]{#1}%
\providecommand \bibfnamefont [1]{#1}%
\providecommand \citenamefont [1]{#1}%
\providecommand \href@noop [0]{\@secondoftwo}%
\providecommand \href [0]{\begingroup \@sanitize@url \@href}%
\providecommand \@href[1]{\@@startlink{#1}\@@href}%
\providecommand \@@href[1]{\endgroup#1\@@endlink}%
\providecommand \@sanitize@url [0]{\catcode `\\12\catcode `\$12\catcode
  `\&12\catcode `\#12\catcode `\^12\catcode `\_12\catcode `\%12\relax}%
\providecommand \@@startlink[1]{}%
\providecommand \@@endlink[0]{}%
\providecommand \url  [0]{\begingroup\@sanitize@url \@url }%
\providecommand \@url [1]{\endgroup\@href {#1}{\urlprefix }}%
\providecommand \urlprefix  [0]{URL }%
\providecommand \Eprint [0]{\href }%
\providecommand \doibase [0]{http://dx.doi.org/}%
\providecommand \selectlanguage [0]{\@gobble}%
\providecommand \bibinfo  [0]{\@secondoftwo}%
\providecommand \bibfield  [0]{\@secondoftwo}%
\providecommand \translation [1]{[#1]}%
\providecommand \BibitemOpen [0]{}%
\providecommand \bibitemStop [0]{}%
\providecommand \bibitemNoStop [0]{.\EOS\space}%
\providecommand \EOS [0]{\spacefactor3000\relax}%
\providecommand \BibitemShut  [1]{\csname bibitem#1\endcsname}%
\let\auto@bib@innerbib\@empty
\bibitem [{\citenamefont {Gibb}\ \emph {et~al.}(2000)\citenamefont {Gibb},
  \citenamefont {Whittet}, \citenamefont {Schutte}, \citenamefont {Boogert},
  \citenamefont {Chiar}, \citenamefont {Ehrenfreund}, \citenamefont
  {Gerakines}, \citenamefont {Keane}, \citenamefont {Tielens}, \citenamefont
  {{van Dishoeck}},\ and\ \citenamefont
  {Kerkhof}}]{gibbInventoryInterstellarIces2000}%
  \BibitemOpen
  \bibfield  {author} {\bibinfo {author} {\bibfnamefont {E.~L.}\ \bibnamefont
  {Gibb}}, \bibinfo {author} {\bibfnamefont {D.~C.~B.}\ \bibnamefont
  {Whittet}}, \bibinfo {author} {\bibfnamefont {W.~A.}\ \bibnamefont
  {Schutte}}, \bibinfo {author} {\bibfnamefont {A.~C.~A.}\ \bibnamefont
  {Boogert}}, \bibinfo {author} {\bibfnamefont {J.~E.}\ \bibnamefont {Chiar}},
  \bibinfo {author} {\bibfnamefont {P.}~\bibnamefont {Ehrenfreund}}, \bibinfo
  {author} {\bibfnamefont {P.~A.}\ \bibnamefont {Gerakines}}, \bibinfo {author}
  {\bibfnamefont {J.~V.}\ \bibnamefont {Keane}}, \bibinfo {author}
  {\bibfnamefont {A.~G. G.~M.}\ \bibnamefont {Tielens}}, \bibinfo {author}
  {\bibfnamefont {E.~F.}\ \bibnamefont {{van Dishoeck}}}, \ and\ \bibinfo
  {author} {\bibfnamefont {O.}~\bibnamefont {Kerkhof}},\ }\href {\doibase
  10.1086/308940} {\bibfield  {journal} {\bibinfo  {journal} {The Astrophysical
  Journal}\ }\textbf {\bibinfo {volume} {536}},\ \bibinfo {pages} {347}
  (\bibinfo {year} {2000})}\BibitemShut {NoStop}%
\bibitem [{\citenamefont {Boogert}, \citenamefont {Gerakines},\ and\
  \citenamefont {Whittet}(2015)}]{boogert2015}%
  \BibitemOpen
  \bibfield  {author} {\bibinfo {author} {\bibfnamefont {A.~A.}\ \bibnamefont
  {Boogert}}, \bibinfo {author} {\bibfnamefont {P.~A.}\ \bibnamefont
  {Gerakines}}, \ and\ \bibinfo {author} {\bibfnamefont {D.~C.}\ \bibnamefont
  {Whittet}},\ }\href {\doibase 10.1146/annurev-astro-082214-122348} {\bibfield
   {journal} {\bibinfo  {journal} {Annual Review of Astronomy and
  Astrophysics}\ }\textbf {\bibinfo {volume} {53}},\ \bibinfo {pages} {541}
  (\bibinfo {year} {2015})}\BibitemShut {NoStop}%
\bibitem [{\citenamefont {McClure}\ \emph {et~al.}(2023)\citenamefont
  {McClure}, \citenamefont {Rocha}, \citenamefont {Pontoppidan}, \citenamefont
  {Crouzet}, \citenamefont {Chu}, \citenamefont {Dartois}, \citenamefont
  {Lamberts}, \citenamefont {Noble}, \citenamefont {Pendleton}, \citenamefont
  {Perotti}, \citenamefont {Qasim}, \citenamefont {Rachid}, \citenamefont
  {Smith}, \citenamefont {Sun}, \citenamefont {Beck}, \citenamefont {Boogert},
  \citenamefont {Brown}, \citenamefont {Caselli}, \citenamefont {Charnley},
  \citenamefont {Cuppen}, \citenamefont {Dickinson}, \citenamefont
  {Drozdovskaya}, \citenamefont {Egami}, \citenamefont {Erkal}, \citenamefont
  {Fraser}, \citenamefont {Garrod}, \citenamefont {Harsono}, \citenamefont
  {Ioppolo}, \citenamefont {{Jim{\'e}nez-Serra}}, \citenamefont {Jin},
  \citenamefont {J{\o}rgensen}, \citenamefont {Kristensen}, \citenamefont
  {Lis}, \citenamefont {McCoustra}, \citenamefont {McGuire}, \citenamefont
  {Melnick}, \citenamefont {{\"O}berg}, \citenamefont {Palumbo}, \citenamefont
  {Shimonishi}, \citenamefont {Sturm}, \citenamefont {Van~Dishoeck},\ and\
  \citenamefont {Linnartz}}]{mcclureIceAgeJWST2023}%
  \BibitemOpen
  \bibfield  {author} {\bibinfo {author} {\bibfnamefont {M.~K.}\ \bibnamefont
  {McClure}}, \bibinfo {author} {\bibfnamefont {W.~R.~M.}\ \bibnamefont
  {Rocha}}, \bibinfo {author} {\bibfnamefont {K.~M.}\ \bibnamefont
  {Pontoppidan}}, \bibinfo {author} {\bibfnamefont {N.}~\bibnamefont
  {Crouzet}}, \bibinfo {author} {\bibfnamefont {L.~E.~U.}\ \bibnamefont {Chu}},
  \bibinfo {author} {\bibfnamefont {E.}~\bibnamefont {Dartois}}, \bibinfo
  {author} {\bibfnamefont {T.}~\bibnamefont {Lamberts}}, \bibinfo {author}
  {\bibfnamefont {J.~A.}\ \bibnamefont {Noble}}, \bibinfo {author}
  {\bibfnamefont {Y.~J.}\ \bibnamefont {Pendleton}}, \bibinfo {author}
  {\bibfnamefont {G.}~\bibnamefont {Perotti}}, \bibinfo {author} {\bibfnamefont
  {D.}~\bibnamefont {Qasim}}, \bibinfo {author} {\bibfnamefont {M.~G.}\
  \bibnamefont {Rachid}}, \bibinfo {author} {\bibfnamefont {Z.~L.}\
  \bibnamefont {Smith}}, \bibinfo {author} {\bibfnamefont {F.}~\bibnamefont
  {Sun}}, \bibinfo {author} {\bibfnamefont {T.~L.}\ \bibnamefont {Beck}},
  \bibinfo {author} {\bibfnamefont {A.~C.~A.}\ \bibnamefont {Boogert}},
  \bibinfo {author} {\bibfnamefont {W.~A.}\ \bibnamefont {Brown}}, \bibinfo
  {author} {\bibfnamefont {P.}~\bibnamefont {Caselli}}, \bibinfo {author}
  {\bibfnamefont {S.~B.}\ \bibnamefont {Charnley}}, \bibinfo {author}
  {\bibfnamefont {H.~M.}\ \bibnamefont {Cuppen}}, \bibinfo {author}
  {\bibfnamefont {H.}~\bibnamefont {Dickinson}}, \bibinfo {author}
  {\bibfnamefont {M.~N.}\ \bibnamefont {Drozdovskaya}}, \bibinfo {author}
  {\bibfnamefont {E.}~\bibnamefont {Egami}}, \bibinfo {author} {\bibfnamefont
  {J.}~\bibnamefont {Erkal}}, \bibinfo {author} {\bibfnamefont
  {H.}~\bibnamefont {Fraser}}, \bibinfo {author} {\bibfnamefont {R.~T.}\
  \bibnamefont {Garrod}}, \bibinfo {author} {\bibfnamefont {D.}~\bibnamefont
  {Harsono}}, \bibinfo {author} {\bibfnamefont {S.}~\bibnamefont {Ioppolo}},
  \bibinfo {author} {\bibfnamefont {I.}~\bibnamefont {{Jim{\'e}nez-Serra}}},
  \bibinfo {author} {\bibfnamefont {M.}~\bibnamefont {Jin}}, \bibinfo {author}
  {\bibfnamefont {J.~K.}\ \bibnamefont {J{\o}rgensen}}, \bibinfo {author}
  {\bibfnamefont {L.~E.}\ \bibnamefont {Kristensen}}, \bibinfo {author}
  {\bibfnamefont {D.~C.}\ \bibnamefont {Lis}}, \bibinfo {author} {\bibfnamefont
  {M.~R.~S.}\ \bibnamefont {McCoustra}}, \bibinfo {author} {\bibfnamefont
  {B.~A.}\ \bibnamefont {McGuire}}, \bibinfo {author} {\bibfnamefont {G.~J.}\
  \bibnamefont {Melnick}}, \bibinfo {author} {\bibfnamefont {K.~I.}\
  \bibnamefont {{\"O}berg}}, \bibinfo {author} {\bibfnamefont {M.~E.}\
  \bibnamefont {Palumbo}}, \bibinfo {author} {\bibfnamefont {T.}~\bibnamefont
  {Shimonishi}}, \bibinfo {author} {\bibfnamefont {J.~A.}\ \bibnamefont
  {Sturm}}, \bibinfo {author} {\bibfnamefont {E.~F.}\ \bibnamefont
  {Van~Dishoeck}}, \ and\ \bibinfo {author} {\bibfnamefont {H.}~\bibnamefont
  {Linnartz}},\ }\href {\doibase 10.1038/s41550-022-01875-w} {\bibfield
  {journal} {\bibinfo  {journal} {Nature Astronomy}\ }\textbf {\bibinfo
  {volume} {7}},\ \bibinfo {pages} {431} (\bibinfo {year} {2023})}\BibitemShut
  {NoStop}%
\bibitem [{\citenamefont {Caselli}\ and\ \citenamefont
  {Ceccarelli}(2012)}]{caselliOurAstrochemicalHeritage2012}%
  \BibitemOpen
  \bibfield  {author} {\bibinfo {author} {\bibfnamefont {P.}~\bibnamefont
  {Caselli}}\ and\ \bibinfo {author} {\bibfnamefont {C.}~\bibnamefont
  {Ceccarelli}},\ }\href {\doibase 10.1007/s00159-012-0056-x} {\bibfield
  {journal} {\bibinfo  {journal} {The Astronomy and Astrophysics Review}\
  }\textbf {\bibinfo {volume} {20}},\ \bibinfo {pages} {56} (\bibinfo {year}
  {2012})}\BibitemShut {NoStop}%
\bibitem [{\citenamefont {Prasad}\ and\ \citenamefont
  {Tarafdar}(1983)}]{prasadUVRadiationField1983}%
  \BibitemOpen
  \bibfield  {author} {\bibinfo {author} {\bibfnamefont {S.~S.}\ \bibnamefont
  {Prasad}}\ and\ \bibinfo {author} {\bibfnamefont {S.~P.}\ \bibnamefont
  {Tarafdar}},\ }\href {\doibase 10.1086/160896} {\bibfield  {journal}
  {\bibinfo  {journal} {The Astrophysical Journal}\ }\textbf {\bibinfo {volume}
  {267}},\ \bibinfo {pages} {603} (\bibinfo {year} {1983})}\BibitemShut
  {NoStop}%
\bibitem [{\citenamefont {Shen}\ \emph {et~al.}(2004)\citenamefont {Shen},
  \citenamefont {Greenberg}, \citenamefont {Schutte},\ and\ \citenamefont
  {Van~Dishoeck}}]{shenCosmicRayInduced2004}%
  \BibitemOpen
  \bibfield  {author} {\bibinfo {author} {\bibfnamefont {C.~J.}\ \bibnamefont
  {Shen}}, \bibinfo {author} {\bibfnamefont {J.~M.}\ \bibnamefont {Greenberg}},
  \bibinfo {author} {\bibfnamefont {W.~A.}\ \bibnamefont {Schutte}}, \ and\
  \bibinfo {author} {\bibfnamefont {E.~F.}\ \bibnamefont {Van~Dishoeck}},\
  }\href {\doibase 10.1051/0004-6361:20031669} {\bibfield  {journal} {\bibinfo
  {journal} {Astronomy \& Astrophysics}\ }\textbf {\bibinfo {volume} {415}},\
  \bibinfo {pages} {203} (\bibinfo {year} {2004})}\BibitemShut {NoStop}%
\bibitem [{\citenamefont {Willacy}\ and\ \citenamefont
  {Langer}(2000)}]{willacy2000}%
  \BibitemOpen
  \bibfield  {author} {\bibinfo {author} {\bibfnamefont {K.}~\bibnamefont
  {Willacy}}\ and\ \bibinfo {author} {\bibfnamefont {W.~D.}\ \bibnamefont
  {Langer}},\ }\href@noop {} {\bibfield  {journal} {\bibinfo  {journal}
  {Astrophysical Journal}\ }\textbf {\bibinfo {volume} {544}},\ \bibinfo
  {pages} {903} (\bibinfo {year} {2000})}\BibitemShut {NoStop}%
\bibitem [{\citenamefont {Pi{\'e}tu}, \citenamefont {Dutrey},\ and\
  \citenamefont {Guilloteau}(2007)}]{pietu2007}%
  \BibitemOpen
  \bibfield  {author} {\bibinfo {author} {\bibfnamefont {V.}~\bibnamefont
  {Pi{\'e}tu}}, \bibinfo {author} {\bibfnamefont {A.}~\bibnamefont {Dutrey}}, \
  and\ \bibinfo {author} {\bibfnamefont {S.}~\bibnamefont {Guilloteau}},\
  }\href {\doibase 10.1051/0004-6361:20066537} {\bibfield  {journal} {\bibinfo
  {journal} {Astronomy \& Astrophysics}\ }\textbf {\bibinfo {volume} {467}},\
  \bibinfo {pages} {163} (\bibinfo {year} {2007})},\ \bibinfo {note} {number: 1
  Publisher: EDP Sciences}\BibitemShut {NoStop}%
\bibitem [{\citenamefont {Dartois}, \citenamefont {Dutrey},\ and\ \citenamefont
  {Guilloteau}(2003)}]{dartoisStructureDMTau}%
  \BibitemOpen
  \bibfield  {author} {\bibinfo {author} {\bibfnamefont {E.}~\bibnamefont
  {Dartois}}, \bibinfo {author} {\bibfnamefont {A.}~\bibnamefont {Dutrey}}, \
  and\ \bibinfo {author} {\bibfnamefont {S.}~\bibnamefont {Guilloteau}},\
  }\href {\doibase 10.1051/0004-6361:20021638} {\bibfield  {journal} {\bibinfo
  {journal} {Astronomy \& Astrophysics}\ }\textbf {\bibinfo {volume} {399}},\
  \bibinfo {pages} {773} (\bibinfo {year} {2003})}\BibitemShut {NoStop}%
\bibitem [{\citenamefont {Dishoeck}\ and\ \citenamefont
  {Black}(1987)}]{dishoeckAbundanceInterstellarCO1987}%
  \BibitemOpen
  \bibfield  {author} {\bibinfo {author} {\bibfnamefont {E.~F.}\ \bibnamefont
  {Dishoeck}}\ and\ \bibinfo {author} {\bibfnamefont {J.~H.}\ \bibnamefont
  {Black}},\ }in\ \href {\doibase 10.1007/978-94-009-3945-5_18} {\emph
  {\bibinfo {booktitle} {Physical {{Processes}} in {{Interstellar Clouds}}}}},\
  \bibinfo {editor} {edited by\ \bibinfo {editor} {\bibfnamefont {G.~E.}\
  \bibnamefont {Morfill}}\ and\ \bibinfo {editor} {\bibfnamefont
  {M.}~\bibnamefont {Scholer}}}\ (\bibinfo  {publisher} {Springer
  Netherlands},\ \bibinfo {address} {Dordrecht},\ \bibinfo {year} {1987})\ pp.\
  \bibinfo {pages} {241--274}\BibitemShut {NoStop}%
\bibitem [{\citenamefont
  {{\"O}berg}(2016)}]{obergPhotochemistryAstrochemistryPhotochemical2016a}%
  \BibitemOpen
  \bibfield  {author} {\bibinfo {author} {\bibfnamefont {K.~I.}\ \bibnamefont
  {{\"O}berg}},\ }\href {\doibase 10.1021/acs.chemrev.5b00694} {\bibfield
  {journal} {\bibinfo  {journal} {Chemical Reviews}\ }\textbf {\bibinfo
  {volume} {116}},\ \bibinfo {pages} {9631} (\bibinfo {year}
  {2016})}\BibitemShut {NoStop}%
\bibitem [{\citenamefont {\"Oberg}\ \emph {et~al.}(2007)\citenamefont
  {\"Oberg}, \citenamefont {Fuchs}, \citenamefont {Awad}, \citenamefont
  {Fraser}, \citenamefont {Schlemmer}, \citenamefont {Van~Dishoeck},\ and\
  \citenamefont {Linnartz}}]{oberg2007}%
  \BibitemOpen
  \bibfield  {author} {\bibinfo {author} {\bibfnamefont {K.~I.}\ \bibnamefont
  {\"Oberg}}, \bibinfo {author} {\bibfnamefont {G.~W.}\ \bibnamefont {Fuchs}},
  \bibinfo {author} {\bibfnamefont {Z.}~\bibnamefont {Awad}}, \bibinfo {author}
  {\bibfnamefont {H.~J.}\ \bibnamefont {Fraser}}, \bibinfo {author}
  {\bibfnamefont {S.}~\bibnamefont {Schlemmer}}, \bibinfo {author}
  {\bibfnamefont {E.~F.}\ \bibnamefont {Van~Dishoeck}}, \ and\ \bibinfo
  {author} {\bibfnamefont {H.}~\bibnamefont {Linnartz}},\ }\href@noop {}
  {\bibfield  {journal} {\bibinfo  {journal} {Astrophysical Journal}\ }\textbf
  {\bibinfo {volume} {662}},\ \bibinfo {pages} {L23} (\bibinfo {year}
  {2007})}\BibitemShut {NoStop}%
\bibitem [{\citenamefont {Mu\~noz Caro}\ \emph {et~al.}(2010)\citenamefont
  {Mu\~noz Caro}, \citenamefont {Jimenez-Escobar}, \citenamefont {Martin-Gago},
  \citenamefont {Rogero}, \citenamefont {Atienza}, \citenamefont {Puertas},
  \citenamefont {Sobrado},\ and\ \citenamefont {Torres-Redondo}}]{munoz2010}%
  \BibitemOpen
  \bibfield  {author} {\bibinfo {author} {\bibfnamefont {G.~M.}\ \bibnamefont
  {Mu\~noz Caro}}, \bibinfo {author} {\bibfnamefont {A.}~\bibnamefont
  {Jimenez-Escobar}}, \bibinfo {author} {\bibfnamefont {J.~A.}\ \bibnamefont
  {Martin-Gago}}, \bibinfo {author} {\bibfnamefont {C.}~\bibnamefont {Rogero}},
  \bibinfo {author} {\bibfnamefont {C.}~\bibnamefont {Atienza}}, \bibinfo
  {author} {\bibfnamefont {S.}~\bibnamefont {Puertas}}, \bibinfo {author}
  {\bibfnamefont {J.~M.}\ \bibnamefont {Sobrado}}, \ and\ \bibinfo {author}
  {\bibfnamefont {J.}~\bibnamefont {Torres-Redondo}},\ }\href@noop {}
  {\bibfield  {journal} {\bibinfo  {journal} {Astronomy \& Astrophysics}\
  }\textbf {\bibinfo {volume} {522}},\ \bibinfo {pages} {A108} (\bibinfo {year}
  {2010})}\BibitemShut {NoStop}%
\bibitem [{\citenamefont {Fayolle}\ \emph {et~al.}(2011)\citenamefont
  {Fayolle}, \citenamefont {Bertin}, \citenamefont {Romanzin}, \citenamefont
  {Michaut}, \citenamefont {\"Oberg}, \citenamefont {Linnartz},\ and\
  \citenamefont {Fillion}}]{fayolle2011}%
  \BibitemOpen
  \bibfield  {author} {\bibinfo {author} {\bibfnamefont {E.~C.}\ \bibnamefont
  {Fayolle}}, \bibinfo {author} {\bibfnamefont {M.}~\bibnamefont {Bertin}},
  \bibinfo {author} {\bibfnamefont {C.}~\bibnamefont {Romanzin}}, \bibinfo
  {author} {\bibfnamefont {X.}~\bibnamefont {Michaut}}, \bibinfo {author}
  {\bibfnamefont {K.~I.}\ \bibnamefont {\"Oberg}}, \bibinfo {author}
  {\bibfnamefont {H.}~\bibnamefont {Linnartz}}, \ and\ \bibinfo {author}
  {\bibfnamefont {J.-H.}\ \bibnamefont {Fillion}},\ }\href@noop {} {\bibfield
  {journal} {\bibinfo  {journal} {Astrophysical Journal Letters}\ }\textbf
  {\bibinfo {volume} {739}},\ \bibinfo {pages} {L36} (\bibinfo {year}
  {2011})}\BibitemShut {NoStop}%
\bibitem [{\citenamefont {Bertin}\ \emph {et~al.}(2012)\citenamefont {Bertin},
  \citenamefont {Fayolle}, \citenamefont {Romanzin}, \citenamefont {\"Oberg},
  \citenamefont {Michaut}, \citenamefont {Moudens}, \citenamefont {Philippe},
  \citenamefont {Jeseck}, \citenamefont {Linnartz},\ and\ \citenamefont
  {Fillion}}]{bertin2012}%
  \BibitemOpen
  \bibfield  {author} {\bibinfo {author} {\bibfnamefont {M.}~\bibnamefont
  {Bertin}}, \bibinfo {author} {\bibfnamefont {E.~C.}\ \bibnamefont {Fayolle}},
  \bibinfo {author} {\bibfnamefont {C.}~\bibnamefont {Romanzin}}, \bibinfo
  {author} {\bibfnamefont {K.~I.}\ \bibnamefont {\"Oberg}}, \bibinfo {author}
  {\bibfnamefont {X.}~\bibnamefont {Michaut}}, \bibinfo {author} {\bibfnamefont
  {A.}~\bibnamefont {Moudens}}, \bibinfo {author} {\bibfnamefont
  {L.}~\bibnamefont {Philippe}}, \bibinfo {author} {\bibfnamefont
  {P.}~\bibnamefont {Jeseck}}, \bibinfo {author} {\bibfnamefont
  {H.}~\bibnamefont {Linnartz}}, \ and\ \bibinfo {author} {\bibfnamefont
  {J.-H.}\ \bibnamefont {Fillion}},\ }\href@noop {} {\bibfield  {journal}
  {\bibinfo  {journal} {Physical Chemistry Chemical Physics}\ }\textbf
  {\bibinfo {volume} {14}},\ \bibinfo {pages} {9929} (\bibinfo {year}
  {2012})}\BibitemShut {NoStop}%
\bibitem [{\citenamefont {Bertin}\ \emph {et~al.}(2013)\citenamefont {Bertin},
  \citenamefont {Fayolle}, \citenamefont {Romanzin}, \citenamefont {Poderoso},
  \citenamefont {Michaut}, \citenamefont {Philippe}, \citenamefont {Jeseck},
  \citenamefont {Öberg}, \citenamefont {Linnartz},\ and\ \citenamefont
  {Fillion}}]{bertin2013}%
  \BibitemOpen
  \bibfield  {author} {\bibinfo {author} {\bibfnamefont {M.}~\bibnamefont
  {Bertin}}, \bibinfo {author} {\bibfnamefont {E.~C.}\ \bibnamefont {Fayolle}},
  \bibinfo {author} {\bibfnamefont {C.}~\bibnamefont {Romanzin}}, \bibinfo
  {author} {\bibfnamefont {H.~A.~M.}\ \bibnamefont {Poderoso}}, \bibinfo
  {author} {\bibfnamefont {X.}~\bibnamefont {Michaut}}, \bibinfo {author}
  {\bibfnamefont {L.}~\bibnamefont {Philippe}}, \bibinfo {author}
  {\bibfnamefont {P.}~\bibnamefont {Jeseck}}, \bibinfo {author} {\bibfnamefont
  {K.~I.}\ \bibnamefont {Öberg}}, \bibinfo {author} {\bibfnamefont
  {H.}~\bibnamefont {Linnartz}}, \ and\ \bibinfo {author} {\bibfnamefont
  {J.-H.}\ \bibnamefont {Fillion}},\ }\href {\doibase
  10.1088/0004-637x/779/2/120} {\bibfield  {journal} {\bibinfo  {journal} {The
  Astrophysical Journal}\ }\textbf {\bibinfo {volume} {779}},\ \bibinfo {pages}
  {120} (\bibinfo {year} {2013})}\BibitemShut {NoStop}%
\bibitem [{\citenamefont {Paardekooper}\ \emph {et~al.}(2016)\citenamefont
  {Paardekooper}, \citenamefont {Fedoseev}, \citenamefont {Riedo},\ and\
  \citenamefont {Linnartz}}]{paardekooper2016}%
  \BibitemOpen
  \bibfield  {author} {\bibinfo {author} {\bibfnamefont {D.~M.}\ \bibnamefont
  {Paardekooper}}, \bibinfo {author} {\bibfnamefont {G.}~\bibnamefont
  {Fedoseev}}, \bibinfo {author} {\bibfnamefont {A.}~\bibnamefont {Riedo}}, \
  and\ \bibinfo {author} {\bibfnamefont {H.}~\bibnamefont {Linnartz}},\ }\href
  {\doibase 10.1051/0004-6361/201629063} {\bibfield  {journal} {\bibinfo
  {journal} {Astronomy \& Astrophysics}\ }\textbf {\bibinfo {volume} {596}},\
  \bibinfo {pages} {A72} (\bibinfo {year} {2016})}\BibitemShut {NoStop}%
\bibitem [{\citenamefont {Carrascosa}\ \emph {et~al.}(2021)\citenamefont
  {Carrascosa}, \citenamefont {Mu{\~n}oz~Caro}, \citenamefont
  {{Gonz{\'a}lez-d{\'i}az}}, \citenamefont {Suevos},\ and\ \citenamefont
  {Chen}}]{carrascosaIntriguingBehaviorUltraviolet2021}%
  \BibitemOpen
  \bibfield  {author} {\bibinfo {author} {\bibfnamefont {H.}~\bibnamefont
  {Carrascosa}}, \bibinfo {author} {\bibfnamefont {G.~M.}\ \bibnamefont
  {Mu{\~n}oz~Caro}}, \bibinfo {author} {\bibfnamefont {C.}~\bibnamefont
  {{Gonz{\'a}lez-d{\'i}az}}}, \bibinfo {author} {\bibfnamefont
  {J.}~\bibnamefont {Suevos}}, \ and\ \bibinfo {author} {\bibfnamefont {Y.-J.}\
  \bibnamefont {Chen}},\ }\href {\doibase 10.3847/1538-4357/ac0a7a} {\bibfield
  {journal} {\bibinfo  {journal} {The Astrophysical Journal}\ }\textbf
  {\bibinfo {volume} {916}},\ \bibinfo {pages} {1} (\bibinfo {year}
  {2021})}\BibitemShut {NoStop}%
\bibitem [{\citenamefont {Gonz{\'a}lez~D{\'i}az}\ \emph
  {et~al.}(2019)\citenamefont {Gonz{\'a}lez~D{\'i}az}, \citenamefont
  {{Carrascosa~de~Lucas}}, \citenamefont {Aparicio}, \citenamefont
  {Mu{\~n}oz~Caro}, \citenamefont {Sie}, \citenamefont {Hsiao}, \citenamefont
  {Cazaux},\ and\ \citenamefont
  {Chen}}]{gonzalezdiazAccretionPhotodesorptionCO2019}%
  \BibitemOpen
  \bibfield  {author} {\bibinfo {author} {\bibfnamefont {C.}~\bibnamefont
  {Gonz{\'a}lez~D{\'i}az}}, \bibinfo {author} {\bibfnamefont {H.}~\bibnamefont
  {{Carrascosa~de~Lucas}}}, \bibinfo {author} {\bibfnamefont {S.}~\bibnamefont
  {Aparicio}}, \bibinfo {author} {\bibfnamefont {G.~M.}\ \bibnamefont
  {Mu{\~n}oz~Caro}}, \bibinfo {author} {\bibfnamefont {N.-E.}\ \bibnamefont
  {Sie}}, \bibinfo {author} {\bibfnamefont {L.-C.}\ \bibnamefont {Hsiao}},
  \bibinfo {author} {\bibfnamefont {S.}~\bibnamefont {Cazaux}}, \ and\ \bibinfo
  {author} {\bibfnamefont {Y.-J.}\ \bibnamefont {Chen}},\ }\href {\doibase
  10.1093/mnras/stz1223} {\bibfield  {journal} {\bibinfo  {journal} {Monthly
  Notices of the Royal Astronomical Society}\ }\textbf {\bibinfo {volume}
  {486}},\ \bibinfo {pages} {5519} (\bibinfo {year} {2019})}\BibitemShut
  {NoStop}%
\bibitem [{\citenamefont {\"Oberg}, \citenamefont {van Dishoeck},\ and\
  \citenamefont {Linnartz}(2009)}]{oberg2009}%
  \BibitemOpen
  \bibfield  {author} {\bibinfo {author} {\bibfnamefont {K.~I.}\ \bibnamefont
  {\"Oberg}}, \bibinfo {author} {\bibfnamefont {E.~F.}\ \bibnamefont {van
  Dishoeck}}, \ and\ \bibinfo {author} {\bibfnamefont {H.}~\bibnamefont
  {Linnartz}},\ }\href@noop {} {\bibfield  {journal} {\bibinfo  {journal}
  {Astronomy \& Astrophysics}\ }\textbf {\bibinfo {volume} {496}},\ \bibinfo
  {pages} {281} (\bibinfo {year} {2009})}\BibitemShut {NoStop}%
\bibitem [{\citenamefont {Sie}\ \emph {et~al.}(2022)\citenamefont {Sie},
  \citenamefont {Cho}, \citenamefont {Huang}, \citenamefont {Caro},
  \citenamefont {Hsiao}, \citenamefont {Lin},\ and\ \citenamefont
  {Chen}}]{sie2022}%
  \BibitemOpen
  \bibfield  {author} {\bibinfo {author} {\bibfnamefont {N.-E.}\ \bibnamefont
  {Sie}}, \bibinfo {author} {\bibfnamefont {Y.-T.}\ \bibnamefont {Cho}},
  \bibinfo {author} {\bibfnamefont {C.-H.}\ \bibnamefont {Huang}}, \bibinfo
  {author} {\bibfnamefont {G.~M.~M.}\ \bibnamefont {Caro}}, \bibinfo {author}
  {\bibfnamefont {L.-C.}\ \bibnamefont {Hsiao}}, \bibinfo {author}
  {\bibfnamefont {H.-C.}\ \bibnamefont {Lin}}, \ and\ \bibinfo {author}
  {\bibfnamefont {Y.-J.}\ \bibnamefont {Chen}},\ }\href {\doibase
  10.3847/1538-4357/ac922a} {\bibfield  {journal} {\bibinfo  {journal} {The
  Astrophysical Journal}\ }\textbf {\bibinfo {volume} {938}},\ \bibinfo {pages}
  {48} (\bibinfo {year} {2022})},\ \bibinfo {note} {publisher: The American
  Astronomical Society}\BibitemShut {NoStop}%
\bibitem [{\citenamefont {Chen}\ \emph {et~al.}(2013)\citenamefont {Chen},
  \citenamefont {Chuang}, \citenamefont {Mu{\~n}oz~Caro}, \citenamefont
  {Nuevo}, \citenamefont {Chu}, \citenamefont {Yih}, \citenamefont {Ip},\ and\
  \citenamefont {Wu}}]{chenVACUUMULTRAVIOLETEMISSION2013}%
  \BibitemOpen
  \bibfield  {author} {\bibinfo {author} {\bibfnamefont {Y.-J.}\ \bibnamefont
  {Chen}}, \bibinfo {author} {\bibfnamefont {K.-J.}\ \bibnamefont {Chuang}},
  \bibinfo {author} {\bibfnamefont {G.~M.}\ \bibnamefont {Mu{\~n}oz~Caro}},
  \bibinfo {author} {\bibfnamefont {M.}~\bibnamefont {Nuevo}}, \bibinfo
  {author} {\bibfnamefont {C.-C.}\ \bibnamefont {Chu}}, \bibinfo {author}
  {\bibfnamefont {T.-S.}\ \bibnamefont {Yih}}, \bibinfo {author} {\bibfnamefont
  {W.-H.}\ \bibnamefont {Ip}}, \ and\ \bibinfo {author} {\bibfnamefont
  {C.-Y.~R.}\ \bibnamefont {Wu}},\ }\href {\doibase 10.1088/0004-637X/781/1/15}
  {\bibfield  {journal} {\bibinfo  {journal} {The Astrophysical Journal}\
  }\textbf {\bibinfo {volume} {781}},\ \bibinfo {pages} {15} (\bibinfo {year}
  {2013})}\BibitemShut {NoStop}%
\bibitem [{\citenamefont {Mu{\~n}oz~Caro}\ \emph {et~al.}(2016)\citenamefont
  {Mu{\~n}oz~Caro}, \citenamefont {Chen}, \citenamefont {Aparicio},
  \citenamefont {Jim{\'e}nez-Escobar}, \citenamefont {Rosu-Finsen},
  \citenamefont {Lasne},\ and\ \citenamefont {McCoustra}}]{munoz2016}%
  \BibitemOpen
  \bibfield  {author} {\bibinfo {author} {\bibfnamefont {G.~M.}\ \bibnamefont
  {Mu{\~n}oz~Caro}}, \bibinfo {author} {\bibfnamefont {Y.-J.}\ \bibnamefont
  {Chen}}, \bibinfo {author} {\bibfnamefont {S.}~\bibnamefont {Aparicio}},
  \bibinfo {author} {\bibfnamefont {A.}~\bibnamefont {Jim{\'e}nez-Escobar}},
  \bibinfo {author} {\bibfnamefont {A.}~\bibnamefont {Rosu-Finsen}}, \bibinfo
  {author} {\bibfnamefont {J.}~\bibnamefont {Lasne}}, \ and\ \bibinfo {author}
  {\bibfnamefont {M.~R.~S.}\ \bibnamefont {McCoustra}},\ }\href {\doibase
  10.1051/0004-6361/201628121} {\bibfield  {journal} {\bibinfo  {journal}
  {Astronomy \& Astrophysics}\ }\textbf {\bibinfo {volume} {589}},\ \bibinfo
  {pages} {A19} (\bibinfo {year} {2016})}\BibitemShut {NoStop}%
\bibitem [{\citenamefont {Lu}\ \emph {et~al.}(2005)\citenamefont {Lu},
  \citenamefont {Chen}, \citenamefont {Cheng}, \citenamefont {Kuo},\ and\
  \citenamefont {Ogilvie}}]{luSpectraVacuumUltraviolet2005}%
  \BibitemOpen
  \bibfield  {author} {\bibinfo {author} {\bibfnamefont {H.-C.}\ \bibnamefont
  {Lu}}, \bibinfo {author} {\bibfnamefont {H.-K.}\ \bibnamefont {Chen}},
  \bibinfo {author} {\bibfnamefont {B.-M.}\ \bibnamefont {Cheng}}, \bibinfo
  {author} {\bibfnamefont {Y.-P.}\ \bibnamefont {Kuo}}, \ and\ \bibinfo
  {author} {\bibfnamefont {J.~F.}\ \bibnamefont {Ogilvie}},\ }\href {\doibase
  10.1088/0953-4075/38/20/006} {\bibfield  {journal} {\bibinfo  {journal}
  {Journal of Physics B: Atomic, Molecular and Optical Physics}\ }\textbf
  {\bibinfo {volume} {38}},\ \bibinfo {pages} {3693} (\bibinfo {year}
  {2005})}\BibitemShut {NoStop}%
\bibitem [{\citenamefont {Del~Fr{\'e}}\ \emph {et~al.}(2023)\citenamefont
  {Del~Fr{\'e}}, \citenamefont {Santamar{\'i}a}, \citenamefont {Duflot},
  \citenamefont {Basalg{\`e}te}, \citenamefont {F{\'e}raud}, \citenamefont
  {Bertin}, \citenamefont {Fillion},\ and\ \citenamefont
  {Monnerville}}]{delfreMechanismUltravioletInducedCO2023}%
  \BibitemOpen
  \bibfield  {author} {\bibinfo {author} {\bibfnamefont {S.}~\bibnamefont
  {Del~Fr{\'e}}}, \bibinfo {author} {\bibfnamefont {A.~R.}\ \bibnamefont
  {Santamar{\'i}a}}, \bibinfo {author} {\bibfnamefont {D.}~\bibnamefont
  {Duflot}}, \bibinfo {author} {\bibfnamefont {R.}~\bibnamefont
  {Basalg{\`e}te}}, \bibinfo {author} {\bibfnamefont {G.}~\bibnamefont
  {F{\'e}raud}}, \bibinfo {author} {\bibfnamefont {M.}~\bibnamefont {Bertin}},
  \bibinfo {author} {\bibfnamefont {J.-H.}\ \bibnamefont {Fillion}}, \ and\
  \bibinfo {author} {\bibfnamefont {M.}~\bibnamefont {Monnerville}},\ }\href
  {\doibase 10.1103/PhysRevLett.131.238001} {\bibfield  {journal} {\bibinfo
  {journal} {Physical Review Letters}\ }\textbf {\bibinfo {volume} {131}},\
  \bibinfo {pages} {238001} (\bibinfo {year} {2023})}\BibitemShut {NoStop}%
\bibitem [{\citenamefont {Hacquard}\ \emph {et~al.}(2024)\citenamefont
  {Hacquard}, \citenamefont {Basalg{\`e}te}, \citenamefont {Del~Fr{\'e}},
  \citenamefont {Rakovsk{\'y}}, \citenamefont {Rivero~Santamaria},
  \citenamefont {Benoit}, \citenamefont {Michaut}, \citenamefont {F{\'e}raud},
  \citenamefont {Bertin}, \citenamefont {Monnerville},\ and\ \citenamefont
  {Fillion}}]{hacquardPhotodesorptionCOIces2024}%
  \BibitemOpen
  \bibfield  {author} {\bibinfo {author} {\bibfnamefont {A.~B.}\ \bibnamefont
  {Hacquard}}, \bibinfo {author} {\bibfnamefont {R.}~\bibnamefont
  {Basalg{\`e}te}}, \bibinfo {author} {\bibfnamefont {S.}~\bibnamefont
  {Del~Fr{\'e}}}, \bibinfo {author} {\bibfnamefont {J.}~\bibnamefont
  {Rakovsk{\'y}}}, \bibinfo {author} {\bibfnamefont {A.}~\bibnamefont
  {Rivero~Santamaria}}, \bibinfo {author} {\bibfnamefont {F.}~\bibnamefont
  {Benoit}}, \bibinfo {author} {\bibfnamefont {X.}~\bibnamefont {Michaut}},
  \bibinfo {author} {\bibfnamefont {G.}~\bibnamefont {F{\'e}raud}}, \bibinfo
  {author} {\bibfnamefont {M.}~\bibnamefont {Bertin}}, \bibinfo {author}
  {\bibfnamefont {M.}~\bibnamefont {Monnerville}}, \ and\ \bibinfo {author}
  {\bibfnamefont {J.-H.}\ \bibnamefont {Fillion}},\ }\href {\doibase
  10.1063/5.0230819} {\bibfield  {journal} {\bibinfo  {journal} {The Journal of
  Chemical Physics}\ }\textbf {\bibinfo {volume} {161}},\ \bibinfo {pages}
  {184306} (\bibinfo {year} {2024})}\BibitemShut {NoStop}%
\bibitem [{\citenamefont {Tokita}\ and\ \citenamefont
  {Behler}(2023)}]{tokitaHowTrainNeural2023}%
  \BibitemOpen
  \bibfield  {author} {\bibinfo {author} {\bibfnamefont {A.~M.}\ \bibnamefont
  {Tokita}}\ and\ \bibinfo {author} {\bibfnamefont {J.}~\bibnamefont
  {Behler}},\ }\href {\doibase 10.1063/5.0160326} {\bibfield  {journal}
  {\bibinfo  {journal} {The Journal of Chemical Physics}\ }\textbf {\bibinfo
  {volume} {159}},\ \bibinfo {pages} {121501} (\bibinfo {year}
  {2023})}\BibitemShut {NoStop}%
\bibitem [{\citenamefont {Unke}\ \emph {et~al.}(2021)\citenamefont {Unke},
  \citenamefont {Chmiela}, \citenamefont {Sauceda}, \citenamefont {Gastegger},
  \citenamefont {Poltavsky}, \citenamefont {Sch{\"u}tt}, \citenamefont
  {Tkatchenko},\ and\ \citenamefont
  {M{\"u}ller}}]{unkeMachineLearningForce2021}%
  \BibitemOpen
  \bibfield  {author} {\bibinfo {author} {\bibfnamefont {O.~T.}\ \bibnamefont
  {Unke}}, \bibinfo {author} {\bibfnamefont {S.}~\bibnamefont {Chmiela}},
  \bibinfo {author} {\bibfnamefont {H.~E.}\ \bibnamefont {Sauceda}}, \bibinfo
  {author} {\bibfnamefont {M.}~\bibnamefont {Gastegger}}, \bibinfo {author}
  {\bibfnamefont {I.}~\bibnamefont {Poltavsky}}, \bibinfo {author}
  {\bibfnamefont {K.~T.}\ \bibnamefont {Sch{\"u}tt}}, \bibinfo {author}
  {\bibfnamefont {A.}~\bibnamefont {Tkatchenko}}, \ and\ \bibinfo {author}
  {\bibfnamefont {K.-R.}\ \bibnamefont {M{\"u}ller}},\ }\href {\doibase
  10.1021/acs.chemrev.0c01111} {\bibfield  {journal} {\bibinfo  {journal}
  {Chemical Reviews}\ }\textbf {\bibinfo {volume} {121}},\ \bibinfo {pages}
  {10142} (\bibinfo {year} {2021})}\BibitemShut {NoStop}%
\bibitem [{\citenamefont {Wang}\ \emph {et~al.}(2024)\citenamefont {Wang},
  \citenamefont {Wang}, \citenamefont {Zhang}, \citenamefont {Li},
  \citenamefont {Zhou},\ and\ \citenamefont
  {Sun}}]{wangMachineLearningInteratomic2024}%
  \BibitemOpen
  \bibfield  {author} {\bibinfo {author} {\bibfnamefont {G.}~\bibnamefont
  {Wang}}, \bibinfo {author} {\bibfnamefont {C.}~\bibnamefont {Wang}}, \bibinfo
  {author} {\bibfnamefont {X.}~\bibnamefont {Zhang}}, \bibinfo {author}
  {\bibfnamefont {Z.}~\bibnamefont {Li}}, \bibinfo {author} {\bibfnamefont
  {J.}~\bibnamefont {Zhou}}, \ and\ \bibinfo {author} {\bibfnamefont
  {Z.}~\bibnamefont {Sun}},\ }\href {\doibase 10.1016/j.isci.2024.109673}
  {\bibfield  {journal} {\bibinfo  {journal} {iScience}\ }\textbf {\bibinfo
  {volume} {27}},\ \bibinfo {pages} {109673} (\bibinfo {year}
  {2024})}\BibitemShut {NoStop}%
\bibitem [{\citenamefont {Rivero~Santamar{\'i}a}\ \emph
  {et~al.}(2021)\citenamefont {Rivero~Santamar{\'i}a}, \citenamefont {Ramos},
  \citenamefont {Alducin}, \citenamefont {Busnengo}, \citenamefont
  {D{\'i}ez~Mui{\~n}o},\ and\ \citenamefont
  {Juaristi}}]{riverosantamariaHighDimensionalAtomisticNeural2021}%
  \BibitemOpen
  \bibfield  {author} {\bibinfo {author} {\bibfnamefont {A.}~\bibnamefont
  {Rivero~Santamar{\'i}a}}, \bibinfo {author} {\bibfnamefont {M.}~\bibnamefont
  {Ramos}}, \bibinfo {author} {\bibfnamefont {M.}~\bibnamefont {Alducin}},
  \bibinfo {author} {\bibfnamefont {H.~F.}\ \bibnamefont {Busnengo}}, \bibinfo
  {author} {\bibfnamefont {R.}~\bibnamefont {D{\'i}ez~Mui{\~n}o}}, \ and\
  \bibinfo {author} {\bibfnamefont {J.~I.}\ \bibnamefont {Juaristi}},\ }\href
  {\doibase 10.1021/acs.jpca.1c00835} {\bibfield  {journal} {\bibinfo
  {journal} {The Journal of Physical Chemistry A}\ }\textbf {\bibinfo {volume}
  {125}},\ \bibinfo {pages} {2588} (\bibinfo {year} {2021})}\BibitemShut
  {NoStop}%
\bibitem [{\citenamefont {Chen}\ \emph {et~al.}(2021)\citenamefont {Chen},
  \citenamefont {Morawietz}, \citenamefont {Mori}, \citenamefont {Markland},\
  and\ \citenamefont {Artrith}}]{chenAENETLAMMPSAENET2021}%
  \BibitemOpen
  \bibfield  {author} {\bibinfo {author} {\bibfnamefont {M.~S.}\ \bibnamefont
  {Chen}}, \bibinfo {author} {\bibfnamefont {T.}~\bibnamefont {Morawietz}},
  \bibinfo {author} {\bibfnamefont {H.}~\bibnamefont {Mori}}, \bibinfo {author}
  {\bibfnamefont {T.~E.}\ \bibnamefont {Markland}}, \ and\ \bibinfo {author}
  {\bibfnamefont {N.}~\bibnamefont {Artrith}},\ }\href {\doibase
  10.1063/5.0063880} {\bibfield  {journal} {\bibinfo  {journal} {The Journal of
  Chemical Physics}\ }\textbf {\bibinfo {volume} {155}},\ \bibinfo {pages}
  {074801} (\bibinfo {year} {2021})}\BibitemShut {NoStop}%
\bibitem [{\citenamefont {F.~Dos~Santos}\ \emph {et~al.}(2025)\citenamefont
  {F.~Dos~Santos}, \citenamefont {Nebgen}, \citenamefont {Allen}, \citenamefont
  {Hamilton}, \citenamefont {Matin}, \citenamefont {Smith},\ and\ \citenamefont
  {Messerly}}]{f.dossantosImprovingBondDissociations2025}%
  \BibitemOpen
  \bibfield  {author} {\bibinfo {author} {\bibfnamefont {L.~G.}\ \bibnamefont
  {F.~Dos~Santos}}, \bibinfo {author} {\bibfnamefont {B.~T.}\ \bibnamefont
  {Nebgen}}, \bibinfo {author} {\bibfnamefont {A.~E.~A.}\ \bibnamefont
  {Allen}}, \bibinfo {author} {\bibfnamefont {B.~W.}\ \bibnamefont {Hamilton}},
  \bibinfo {author} {\bibfnamefont {S.}~\bibnamefont {Matin}}, \bibinfo
  {author} {\bibfnamefont {J.~S.}\ \bibnamefont {Smith}}, \ and\ \bibinfo
  {author} {\bibfnamefont {R.~A.}\ \bibnamefont {Messerly}},\ }\href {\doibase
  10.1021/acs.jcim.4c01847} {\bibfield  {journal} {\bibinfo  {journal} {Journal
  of Chemical Information and Modeling}\ }\textbf {\bibinfo {volume} {65}},\
  \bibinfo {pages} {1198} (\bibinfo {year} {2025})}\BibitemShut {NoStop}%
\bibitem [{\citenamefont {Wang}\ \emph {et~al.}(2018)\citenamefont {Wang},
  \citenamefont {Zhang}, \citenamefont {Han},\ and\ \citenamefont
  {E}}]{WangDeePMDkitdeeplearning2018}%
  \BibitemOpen
  \bibfield  {author} {\bibinfo {author} {\bibfnamefont {H.}~\bibnamefont
  {Wang}}, \bibinfo {author} {\bibfnamefont {L.}~\bibnamefont {Zhang}},
  \bibinfo {author} {\bibfnamefont {J.}~\bibnamefont {Han}}, \ and\ \bibinfo
  {author} {\bibfnamefont {W.}~\bibnamefont {E}},\ }\href {\doibase
  10.1016/j.cpc.2018.03.016} {\bibfield  {journal} {\bibinfo  {journal}
  {Computer Physics Communications}\ }\textbf {\bibinfo {volume} {228}},\
  \bibinfo {pages} {178} (\bibinfo {year} {2018})}\BibitemShut {NoStop}%
\bibitem [{\citenamefont {Zeng}\ \emph {et~al.}(2023)\citenamefont {Zeng},
  \citenamefont {Zhang}, \citenamefont {Lu}, \citenamefont {Mo}, \citenamefont
  {Li}, \citenamefont {Chen}, \citenamefont {Rynik}, \citenamefont {Huang},
  \citenamefont {Li}, \citenamefont {Shi}, \citenamefont {Wang}, \citenamefont
  {Ye}, \citenamefont {Tuo}, \citenamefont {Yang}, \citenamefont {Ding},
  \citenamefont {Li}, \citenamefont {Tisi}, \citenamefont {Zeng}, \citenamefont
  {Bao}, \citenamefont {Xia}, \citenamefont {Huang}, \citenamefont {Muraoka},
  \citenamefont {Wang}, \citenamefont {Chang}, \citenamefont {Yuan},
  \citenamefont {Bore}, \citenamefont {Cai}, \citenamefont {Lin}, \citenamefont
  {Wang}, \citenamefont {Xu}, \citenamefont {Zhu}, \citenamefont {Luo},
  \citenamefont {Zhang}, \citenamefont {Goodall}, \citenamefont {Liang},
  \citenamefont {Singh}, \citenamefont {Yao}, \citenamefont {Zhang},
  \citenamefont {Wentzcovitch}, \citenamefont {Han}, \citenamefont {Liu},
  \citenamefont {Jia}, \citenamefont {York}, \citenamefont {E}, \citenamefont
  {Car}, \citenamefont {Zhang},\ and\ \citenamefont
  {Wang}}]{ZengDeePMDkitv2software2023}%
  \BibitemOpen
  \bibfield  {author} {\bibinfo {author} {\bibfnamefont {J.}~\bibnamefont
  {Zeng}}, \bibinfo {author} {\bibfnamefont {D.}~\bibnamefont {Zhang}},
  \bibinfo {author} {\bibfnamefont {D.}~\bibnamefont {Lu}}, \bibinfo {author}
  {\bibfnamefont {P.}~\bibnamefont {Mo}}, \bibinfo {author} {\bibfnamefont
  {Z.}~\bibnamefont {Li}}, \bibinfo {author} {\bibfnamefont {Y.}~\bibnamefont
  {Chen}}, \bibinfo {author} {\bibfnamefont {M.}~\bibnamefont {Rynik}},
  \bibinfo {author} {\bibfnamefont {L.}~\bibnamefont {Huang}}, \bibinfo
  {author} {\bibfnamefont {Z.}~\bibnamefont {Li}}, \bibinfo {author}
  {\bibfnamefont {S.}~\bibnamefont {Shi}}, \bibinfo {author} {\bibfnamefont
  {Y.}~\bibnamefont {Wang}}, \bibinfo {author} {\bibfnamefont {H.}~\bibnamefont
  {Ye}}, \bibinfo {author} {\bibfnamefont {P.}~\bibnamefont {Tuo}}, \bibinfo
  {author} {\bibfnamefont {J.}~\bibnamefont {Yang}}, \bibinfo {author}
  {\bibfnamefont {Y.}~\bibnamefont {Ding}}, \bibinfo {author} {\bibfnamefont
  {Y.}~\bibnamefont {Li}}, \bibinfo {author} {\bibfnamefont {D.}~\bibnamefont
  {Tisi}}, \bibinfo {author} {\bibfnamefont {Q.}~\bibnamefont {Zeng}}, \bibinfo
  {author} {\bibfnamefont {H.}~\bibnamefont {Bao}}, \bibinfo {author}
  {\bibfnamefont {Y.}~\bibnamefont {Xia}}, \bibinfo {author} {\bibfnamefont
  {J.}~\bibnamefont {Huang}}, \bibinfo {author} {\bibfnamefont
  {K.}~\bibnamefont {Muraoka}}, \bibinfo {author} {\bibfnamefont
  {Y.}~\bibnamefont {Wang}}, \bibinfo {author} {\bibfnamefont {J.}~\bibnamefont
  {Chang}}, \bibinfo {author} {\bibfnamefont {F.}~\bibnamefont {Yuan}},
  \bibinfo {author} {\bibfnamefont {S.~L.}\ \bibnamefont {Bore}}, \bibinfo
  {author} {\bibfnamefont {C.}~\bibnamefont {Cai}}, \bibinfo {author}
  {\bibfnamefont {Y.}~\bibnamefont {Lin}}, \bibinfo {author} {\bibfnamefont
  {B.}~\bibnamefont {Wang}}, \bibinfo {author} {\bibfnamefont {J.}~\bibnamefont
  {Xu}}, \bibinfo {author} {\bibfnamefont {J.-X.}\ \bibnamefont {Zhu}},
  \bibinfo {author} {\bibfnamefont {C.}~\bibnamefont {Luo}}, \bibinfo {author}
  {\bibfnamefont {Y.}~\bibnamefont {Zhang}}, \bibinfo {author} {\bibfnamefont
  {R.~E.~A.}\ \bibnamefont {Goodall}}, \bibinfo {author} {\bibfnamefont
  {W.}~\bibnamefont {Liang}}, \bibinfo {author} {\bibfnamefont {A.~K.}\
  \bibnamefont {Singh}}, \bibinfo {author} {\bibfnamefont {S.}~\bibnamefont
  {Yao}}, \bibinfo {author} {\bibfnamefont {J.}~\bibnamefont {Zhang}}, \bibinfo
  {author} {\bibfnamefont {R.}~\bibnamefont {Wentzcovitch}}, \bibinfo {author}
  {\bibfnamefont {J.}~\bibnamefont {Han}}, \bibinfo {author} {\bibfnamefont
  {J.}~\bibnamefont {Liu}}, \bibinfo {author} {\bibfnamefont {W.}~\bibnamefont
  {Jia}}, \bibinfo {author} {\bibfnamefont {D.~M.}\ \bibnamefont {York}},
  \bibinfo {author} {\bibfnamefont {W.}~\bibnamefont {E}}, \bibinfo {author}
  {\bibfnamefont {R.}~\bibnamefont {Car}}, \bibinfo {author} {\bibfnamefont
  {L.}~\bibnamefont {Zhang}}, \ and\ \bibinfo {author} {\bibfnamefont
  {H.}~\bibnamefont {Wang}},\ }\href {\doibase 10.1063/5.0155600} {\bibfield
  {journal} {\bibinfo  {journal} {The Journal of Chemical Physics}\ }\textbf
  {\bibinfo {volume} {159}},\ \bibinfo {pages} {054801} (\bibinfo {year}
  {2023})}\BibitemShut {NoStop}%
\bibitem [{\citenamefont {Behler}\ and\ \citenamefont
  {Parrinello}(2007{\natexlab{a}})}]{PhysRevLett_98_146401}%
  \BibitemOpen
  \bibfield  {author} {\bibinfo {author} {\bibfnamefont {J.}~\bibnamefont
  {Behler}}\ and\ \bibinfo {author} {\bibfnamefont {M.}~\bibnamefont
  {Parrinello}},\ }\href {\doibase 10.1103/PhysRevLett.98.146401} {\bibfield
  {journal} {\bibinfo  {journal} {Phys. Rev. Lett.}\ }\textbf {\bibinfo
  {volume} {98}},\ \bibinfo {pages} {146401} (\bibinfo {year}
  {2007}{\natexlab{a}})}\BibitemShut {NoStop}%
\bibitem [{\citenamefont {Wen}\ \emph {et~al.}(2022)\citenamefont {Wen},
  \citenamefont {Zhang}, \citenamefont {Wang}, \citenamefont {E},\ and\
  \citenamefont {Srolovitz}}]{WenDeeppotentialsmaterials2022}%
  \BibitemOpen
  \bibfield  {author} {\bibinfo {author} {\bibfnamefont {T.}~\bibnamefont
  {Wen}}, \bibinfo {author} {\bibfnamefont {L.}~\bibnamefont {Zhang}}, \bibinfo
  {author} {\bibfnamefont {H.}~\bibnamefont {Wang}}, \bibinfo {author}
  {\bibfnamefont {W.}~\bibnamefont {E}}, \ and\ \bibinfo {author}
  {\bibfnamefont {D.~J.}\ \bibnamefont {Srolovitz}},\ }\href {\doibase
  10.1088/2752-5724/ac681d} {\bibfield  {journal} {\bibinfo  {journal}
  {Materials Futures}\ }\textbf {\bibinfo {volume} {1}},\ \bibinfo {pages}
  {022601} (\bibinfo {year} {2022})}\BibitemShut {NoStop}%
\bibitem [{\citenamefont {Kingma}(2014)}]{kingma2014adam}%
  \BibitemOpen
  \bibfield  {author} {\bibinfo {author} {\bibfnamefont {D.~P.}\ \bibnamefont
  {Kingma}},\ }\href@noop {} {\bibfield  {journal} {\bibinfo  {journal} {arXiv
  preprint arXiv:1412.6980}\ } (\bibinfo {year} {2014})}\BibitemShut {NoStop}%
\bibitem [{\citenamefont {Kresse}\ and\ \citenamefont
  {Furthm{\"u}ller}(1996{\natexlab{a}})}]{Kresse_96_1}%
  \BibitemOpen
  \bibfield  {author} {\bibinfo {author} {\bibfnamefont {G.}~\bibnamefont
  {Kresse}}\ and\ \bibinfo {author} {\bibfnamefont {J.}~\bibnamefont
  {Furthm{\"u}ller}},\ }\href {\doibase 10.1016/0927-0256(96)00008-0}
  {\bibfield  {journal} {\bibinfo  {journal} {Comput. Mater. Sci.}\ }\textbf
  {\bibinfo {volume} {6}},\ \bibinfo {pages} {15} (\bibinfo {year}
  {1996}{\natexlab{a}})}\BibitemShut {NoStop}%
\bibitem [{\citenamefont {Kresse}\ and\ \citenamefont
  {Furthm{\"u}ller}(1996{\natexlab{b}})}]{Kresse_96_2}%
  \BibitemOpen
  \bibfield  {author} {\bibinfo {author} {\bibfnamefont {G.}~\bibnamefont
  {Kresse}}\ and\ \bibinfo {author} {\bibfnamefont {J.}~\bibnamefont
  {Furthm{\"u}ller}},\ }\href {\doibase 10.1103/physrevb.54.11169} {\bibfield
  {journal} {\bibinfo  {journal} {Physical Review B}\ }\textbf {\bibinfo
  {volume} {54}},\ \bibinfo {pages} {11169} (\bibinfo {year}
  {1996}{\natexlab{b}})}\BibitemShut {NoStop}%
\bibitem [{\citenamefont {Perdew}, \citenamefont {Burke},\ and\ \citenamefont
  {Ernzerhof}(1996)}]{Perdew_96}%
  \BibitemOpen
  \bibfield  {author} {\bibinfo {author} {\bibfnamefont {J.~P.}\ \bibnamefont
  {Perdew}}, \bibinfo {author} {\bibfnamefont {K.}~\bibnamefont {Burke}}, \
  and\ \bibinfo {author} {\bibfnamefont {M.}~\bibnamefont {Ernzerhof}},\ }\href
  {\doibase 10.1103/physrevlett.77.3865} {\bibfield  {journal} {\bibinfo
  {journal} {Physical Review Letters}\ }\textbf {\bibinfo {volume} {77}},\
  \bibinfo {pages} {3865} (\bibinfo {year} {1996})}\BibitemShut {NoStop}%
\bibitem [{\citenamefont {Grimme}\ \emph {et~al.}(2010)\citenamefont {Grimme},
  \citenamefont {Antony}, \citenamefont {Ehrlich},\ and\ \citenamefont
  {Krieg}}]{Grimme_2010}%
  \BibitemOpen
  \bibfield  {author} {\bibinfo {author} {\bibfnamefont {S.}~\bibnamefont
  {Grimme}}, \bibinfo {author} {\bibfnamefont {J.}~\bibnamefont {Antony}},
  \bibinfo {author} {\bibfnamefont {S.}~\bibnamefont {Ehrlich}}, \ and\
  \bibinfo {author} {\bibfnamefont {H.}~\bibnamefont {Krieg}},\ }\href
  {\doibase 10.1063/1.3382344} {\bibfield  {journal} {\bibinfo  {journal} {J.
  Chem. Phys.}\ }\textbf {\bibinfo {volume} {132}},\ \bibinfo {pages} {154104}
  (\bibinfo {year} {2010})}\BibitemShut {NoStop}%
\bibitem [{\citenamefont {Grimme}, \citenamefont {Ehrlich},\ and\ \citenamefont
  {Goerigk}(2011)}]{Grimme_2011}%
  \BibitemOpen
  \bibfield  {author} {\bibinfo {author} {\bibfnamefont {S.}~\bibnamefont
  {Grimme}}, \bibinfo {author} {\bibfnamefont {S.}~\bibnamefont {Ehrlich}}, \
  and\ \bibinfo {author} {\bibfnamefont {L.}~\bibnamefont {Goerigk}},\ }\href
  {\doibase 10.1002/jcc.21759} {\bibfield  {journal} {\bibinfo  {journal} {J.
  Comput. Chem.}\ }\textbf {\bibinfo {volume} {32}},\ \bibinfo {pages} {1456}
  (\bibinfo {year} {2011})}\BibitemShut {NoStop}%
\bibitem [{\citenamefont {Bl{\"o}chl}(1994)}]{bl1994p}%
  \BibitemOpen
  \bibfield  {author} {\bibinfo {author} {\bibfnamefont {P.~E.}\ \bibnamefont
  {Bl{\"o}chl}},\ }\href@noop {} {\bibfield  {journal} {\bibinfo  {journal}
  {Phys Rev B}\ }\textbf {\bibinfo {volume} {50}},\ \bibinfo {pages} {17953}
  (\bibinfo {year} {1994})}\BibitemShut {NoStop}%
\bibitem [{\citenamefont {Kresse}\ and\ \citenamefont
  {Joubert}(1999)}]{Kresseultrasoftpseudopotentialsprojector1999a}%
  \BibitemOpen
  \bibfield  {author} {\bibinfo {author} {\bibfnamefont {G.}~\bibnamefont
  {Kresse}}\ and\ \bibinfo {author} {\bibfnamefont {D.}~\bibnamefont
  {Joubert}},\ }\href {\doibase 10.1103/PhysRevB.59.1758} {\bibfield  {journal}
  {\bibinfo  {journal} {Physical Review B}\ }\textbf {\bibinfo {volume} {59}},\
  \bibinfo {pages} {1758} (\bibinfo {year} {1999})}\BibitemShut {NoStop}%
\bibitem [{\citenamefont {Plimpton}\ \emph {et~al.}(2023)\citenamefont
  {Plimpton}, \citenamefont {Kohlmeyer}, \citenamefont {Thompson},
  \citenamefont {Moore},\ and\ \citenamefont
  {Berger}}]{plimpton_2023_10806852}%
  \BibitemOpen
  \bibfield  {author} {\bibinfo {author} {\bibfnamefont {S.~J.}\ \bibnamefont
  {Plimpton}}, \bibinfo {author} {\bibfnamefont {A.}~\bibnamefont {Kohlmeyer}},
  \bibinfo {author} {\bibfnamefont {A.~P.}\ \bibnamefont {Thompson}}, \bibinfo
  {author} {\bibfnamefont {S.~G.}\ \bibnamefont {Moore}}, \ and\ \bibinfo
  {author} {\bibfnamefont {R.}~\bibnamefont {Berger}},\ }\href {\doibase
  10.5281/zenodo.10806852} {\enquote {\bibinfo {title} {Lammps: Large-scale
  atomic/molecular massively parallel simulator},}\ } (\bibinfo {year}
  {2023})\BibitemShut {NoStop}%
\bibitem [{\citenamefont
  {Nos{\'e}}(1984)}]{Noseunifiedformulationconstant1984}%
  \BibitemOpen
  \bibfield  {author} {\bibinfo {author} {\bibfnamefont {S.}~\bibnamefont
  {Nos{\'e}}},\ }\href {\doibase 10.1063/1.447334} {\bibfield  {journal}
  {\bibinfo  {journal} {The Journal of Chemical Physics}\ }\textbf {\bibinfo
  {volume} {81}},\ \bibinfo {pages} {511} (\bibinfo {year} {1984})}\BibitemShut
  {NoStop}%
\bibitem [{\citenamefont
  {Hoover}(1985)}]{HooverCanonicaldynamicsEquilibrium1985}%
  \BibitemOpen
  \bibfield  {author} {\bibinfo {author} {\bibfnamefont {W.~G.}\ \bibnamefont
  {Hoover}},\ }\href {\doibase 10.1103/PhysRevA.31.1695} {\bibfield  {journal}
  {\bibinfo  {journal} {Physical Review A}\ }\textbf {\bibinfo {volume} {31}},\
  \bibinfo {pages} {1695} (\bibinfo {year} {1985})}\BibitemShut {NoStop}%
\bibitem [{\citenamefont {Verlet}(1967)}]{VerletComputerExperiment1967}%
  \BibitemOpen
  \bibfield  {author} {\bibinfo {author} {\bibfnamefont {L.}~\bibnamefont
  {Verlet}},\ }\href {\doibase 10.1103/PhysRev.159.98} {\bibfield  {journal}
  {\bibinfo  {journal} {Phys. Rev.}\ }\textbf {\bibinfo {volume} {159}},\
  \bibinfo {pages} {98} (\bibinfo {year} {1967})}\BibitemShut {NoStop}%
\bibitem [{\citenamefont {Swope}\ \emph {et~al.}(1982)\citenamefont {Swope},
  \citenamefont {Andersen}, \citenamefont {Berens},\ and\ \citenamefont
  {Wilson}}]{SwopeComputerSimulationMethod1982}%
  \BibitemOpen
  \bibfield  {author} {\bibinfo {author} {\bibfnamefont {W.~C.}\ \bibnamefont
  {Swope}}, \bibinfo {author} {\bibfnamefont {H.~C.}\ \bibnamefont {Andersen}},
  \bibinfo {author} {\bibfnamefont {P.~H.}\ \bibnamefont {Berens}}, \ and\
  \bibinfo {author} {\bibfnamefont {K.~R.}\ \bibnamefont {Wilson}},\ }\href
  {\doibase 10.1063/1.442716} {\bibfield  {journal} {\bibinfo  {journal} {The
  Journal of Chemical Physics}\ }\textbf {\bibinfo {volume} {76}},\ \bibinfo
  {pages} {637} (\bibinfo {year} {1982})},\ \Eprint
  {http://arxiv.org/abs/https://pubs.aip.org/aip/jcp/article-pdf/76/1/637/18934474/637\_1\_online.pdf}
  {https://pubs.aip.org/aip/jcp/article-pdf/76/1/637/18934474/637\_1\_online.pdf}
  \BibitemShut {NoStop}%
\bibitem [{\citenamefont {Billing}(2000)}]{billing2000BooK}%
  \BibitemOpen
  \bibfield  {author} {\bibinfo {author} {\bibfnamefont {G.~D.}\ \bibnamefont
  {Billing}},\ }\href@noop {} {\emph {\bibinfo {title} {Dynamics of Molecule
  Surface Interaction}}}\ (\bibinfo  {publisher} {John Wiley \& Sons},\
  \bibinfo {year} {2000})\BibitemShut {NoStop}%
\bibitem [{\citenamefont {Behler}\ and\ \citenamefont
  {Parrinello}(2007{\natexlab{b}})}]{BehlerGeneralizedNeuralNetworkRepresentation2007}%
  \BibitemOpen
  \bibfield  {author} {\bibinfo {author} {\bibfnamefont {J.}~\bibnamefont
  {Behler}}\ and\ \bibinfo {author} {\bibfnamefont {M.}~\bibnamefont
  {Parrinello}},\ }\href {\doibase 10.1103/PhysRevLett.98.146401} {\bibfield
  {journal} {\bibinfo  {journal} {Physical Review Letters}\ }\textbf {\bibinfo
  {volume} {98}},\ \bibinfo {pages} {146401} (\bibinfo {year}
  {2007}{\natexlab{b}})}\BibitemShut {NoStop}%
\bibitem [{\citenamefont
  {Behler}(2011)}]{BehlerAtomcenteredsymmetryfunctions2011}%
  \BibitemOpen
  \bibfield  {author} {\bibinfo {author} {\bibfnamefont {J.}~\bibnamefont
  {Behler}},\ }\href {\doibase 10.1063/1.3553717} {\bibfield  {journal}
  {\bibinfo  {journal} {The Journal of Chemical Physics}\ }\textbf {\bibinfo
  {volume} {134}},\ \bibinfo {pages} {074106} (\bibinfo {year}
  {2011})}\BibitemShut {NoStop}%
\bibitem [{\citenamefont {Hedman}\ \emph {et~al.}(2024)\citenamefont {Hedman},
  \citenamefont {McLean}, \citenamefont {Bichara}, \citenamefont {Maruyama},
  \citenamefont {Larsson},\ and\ \citenamefont
  {Ding}}]{HedmanDynamicsgrowingcarbon2024}%
  \BibitemOpen
  \bibfield  {author} {\bibinfo {author} {\bibfnamefont {D.}~\bibnamefont
  {Hedman}}, \bibinfo {author} {\bibfnamefont {B.}~\bibnamefont {McLean}},
  \bibinfo {author} {\bibfnamefont {C.}~\bibnamefont {Bichara}}, \bibinfo
  {author} {\bibfnamefont {S.}~\bibnamefont {Maruyama}}, \bibinfo {author}
  {\bibfnamefont {J.~A.}\ \bibnamefont {Larsson}}, \ and\ \bibinfo {author}
  {\bibfnamefont {F.}~\bibnamefont {Ding}},\ }\href {\doibase
  10.1038/s41467-024-47999-7} {\bibfield  {journal} {\bibinfo  {journal}
  {Nature Communications}\ }\textbf {\bibinfo {volume} {15}},\ \bibinfo {pages}
  {4076} (\bibinfo {year} {2024})}\BibitemShut {NoStop}%
\bibitem [{\citenamefont {Zhang}\ \emph {et~al.}(2024)\citenamefont {Zhang},
  \citenamefont {Wang}, \citenamefont {Liu}, \citenamefont {Fan}, \citenamefont
  {Cao},\ and\ \citenamefont {Xiao}}]{ZhangPolarizationdrivenbandtopology2024}%
  \BibitemOpen
  \bibfield  {author} {\bibinfo {author} {\bibfnamefont {X.-W.}\ \bibnamefont
  {Zhang}}, \bibinfo {author} {\bibfnamefont {C.}~\bibnamefont {Wang}},
  \bibinfo {author} {\bibfnamefont {X.}~\bibnamefont {Liu}}, \bibinfo {author}
  {\bibfnamefont {Y.}~\bibnamefont {Fan}}, \bibinfo {author} {\bibfnamefont
  {T.}~\bibnamefont {Cao}}, \ and\ \bibinfo {author} {\bibfnamefont
  {D.}~\bibnamefont {Xiao}},\ }\href {\doibase 10.1038/s41467-024-48511-x}
  {\bibfield  {journal} {\bibinfo  {journal} {Nature Communications}\ }\textbf
  {\bibinfo {volume} {15}},\ \bibinfo {pages} {4223} (\bibinfo {year}
  {2024})}\BibitemShut {NoStop}%
\bibitem [{\citenamefont {Tosello~Gardini}, \citenamefont {Raucci},\ and\
  \citenamefont
  {Parrinello}(2025)}]{ToselloGardiniMachinelearningdrivenmolecular2025}%
  \BibitemOpen
  \bibfield  {author} {\bibinfo {author} {\bibfnamefont {A.}~\bibnamefont
  {Tosello~Gardini}}, \bibinfo {author} {\bibfnamefont {U.}~\bibnamefont
  {Raucci}}, \ and\ \bibinfo {author} {\bibfnamefont {M.}~\bibnamefont
  {Parrinello}},\ }\href {\doibase 10.1038/s41467-025-57688-8} {\bibfield
  {journal} {\bibinfo  {journal} {Nature Communications}\ }\textbf {\bibinfo
  {volume} {16}},\ \bibinfo {pages} {2475} (\bibinfo {year}
  {2025})}\BibitemShut {NoStop}%
\bibitem [{\citenamefont {Virtanen}\ \emph {et~al.}(2020)\citenamefont
  {Virtanen}, \citenamefont {Gommers}, \citenamefont {Oliphant}, \citenamefont
  {Haberland}, \citenamefont {Reddy}, \citenamefont {Cournapeau}, \citenamefont
  {Burovski}, \citenamefont {Peterson}, \citenamefont {Weckesser},
  \citenamefont {Bright}, \citenamefont {{van der Walt}}, \citenamefont
  {Brett}, \citenamefont {Wilson}, \citenamefont {Millman}, \citenamefont
  {Mayorov}, \citenamefont {Nelson}, \citenamefont {Jones}, \citenamefont
  {Kern}, \citenamefont {Larson}, \citenamefont {Carey}, \citenamefont {Polat},
  \citenamefont {Feng}, \citenamefont {Moore}, \citenamefont {{VanderPlas}},
  \citenamefont {Laxalde}, \citenamefont {Perktold}, \citenamefont {Cimrman},
  \citenamefont {Henriksen}, \citenamefont {Quintero}, \citenamefont {Harris},
  \citenamefont {Archibald}, \citenamefont {Ribeiro}, \citenamefont
  {Pedregosa}, \citenamefont {{van Mulbregt}},\ and\ \citenamefont {{SciPy 1.0
  Contributors}}}]{2020SciPy-NMeth}%
  \BibitemOpen
  \bibfield  {author} {\bibinfo {author} {\bibfnamefont {P.}~\bibnamefont
  {Virtanen}}, \bibinfo {author} {\bibfnamefont {R.}~\bibnamefont {Gommers}},
  \bibinfo {author} {\bibfnamefont {T.~E.}\ \bibnamefont {Oliphant}}, \bibinfo
  {author} {\bibfnamefont {M.}~\bibnamefont {Haberland}}, \bibinfo {author}
  {\bibfnamefont {T.}~\bibnamefont {Reddy}}, \bibinfo {author} {\bibfnamefont
  {D.}~\bibnamefont {Cournapeau}}, \bibinfo {author} {\bibfnamefont
  {E.}~\bibnamefont {Burovski}}, \bibinfo {author} {\bibfnamefont
  {P.}~\bibnamefont {Peterson}}, \bibinfo {author} {\bibfnamefont
  {W.}~\bibnamefont {Weckesser}}, \bibinfo {author} {\bibfnamefont
  {J.}~\bibnamefont {Bright}}, \bibinfo {author} {\bibfnamefont {S.~J.}\
  \bibnamefont {{van der Walt}}}, \bibinfo {author} {\bibfnamefont
  {M.}~\bibnamefont {Brett}}, \bibinfo {author} {\bibfnamefont
  {J.}~\bibnamefont {Wilson}}, \bibinfo {author} {\bibfnamefont {K.~J.}\
  \bibnamefont {Millman}}, \bibinfo {author} {\bibfnamefont {N.}~\bibnamefont
  {Mayorov}}, \bibinfo {author} {\bibfnamefont {A.~R.~J.}\ \bibnamefont
  {Nelson}}, \bibinfo {author} {\bibfnamefont {E.}~\bibnamefont {Jones}},
  \bibinfo {author} {\bibfnamefont {R.}~\bibnamefont {Kern}}, \bibinfo {author}
  {\bibfnamefont {E.}~\bibnamefont {Larson}}, \bibinfo {author} {\bibfnamefont
  {C.~J.}\ \bibnamefont {Carey}}, \bibinfo {author} {\bibfnamefont
  {{\.I}.}~\bibnamefont {Polat}}, \bibinfo {author} {\bibfnamefont
  {Y.}~\bibnamefont {Feng}}, \bibinfo {author} {\bibfnamefont {E.~W.}\
  \bibnamefont {Moore}}, \bibinfo {author} {\bibfnamefont {J.}~\bibnamefont
  {{VanderPlas}}}, \bibinfo {author} {\bibfnamefont {D.}~\bibnamefont
  {Laxalde}}, \bibinfo {author} {\bibfnamefont {J.}~\bibnamefont {Perktold}},
  \bibinfo {author} {\bibfnamefont {R.}~\bibnamefont {Cimrman}}, \bibinfo
  {author} {\bibfnamefont {I.}~\bibnamefont {Henriksen}}, \bibinfo {author}
  {\bibfnamefont {E.~A.}\ \bibnamefont {Quintero}}, \bibinfo {author}
  {\bibfnamefont {C.~R.}\ \bibnamefont {Harris}}, \bibinfo {author}
  {\bibfnamefont {A.~M.}\ \bibnamefont {Archibald}}, \bibinfo {author}
  {\bibfnamefont {A.~H.}\ \bibnamefont {Ribeiro}}, \bibinfo {author}
  {\bibfnamefont {F.}~\bibnamefont {Pedregosa}}, \bibinfo {author}
  {\bibfnamefont {P.}~\bibnamefont {{van Mulbregt}}}, \ and\ \bibinfo {author}
  {\bibnamefont {{SciPy 1.0 Contributors}}},\ }\href {\doibase
  10.1038/s41592-019-0686-2} {\bibfield  {journal} {\bibinfo  {journal} {Nature
  Methods}\ }\textbf {\bibinfo {volume} {17}},\ \bibinfo {pages} {261}
  (\bibinfo {year} {2020})}\BibitemShut {NoStop}%
\end{thebibliography}%

\end{document}